\documentclass[aps,prb,longbibliography,showpacs,floatfix,superscriptaddress,twocolumn]{revtex4-1}

\usepackage{graphicx,color}
\usepackage{amsfonts}
\usepackage[figuresright]{rotating}
\usepackage{amssymb}
\usepackage{amsmath}
\usepackage{mathtools}
\usepackage{psfrag}
\usepackage{subfigure}
\usepackage{multirow}
\usepackage{tabularx}
\usepackage{textcomp}
\usepackage{units}
\usepackage{lipsum}
\usepackage{soul}
\usepackage{titlesec}
\usepackage{times}
\usepackage{hyperref}
\usepackage{enumitem}

\DeclareMathAlphabet\mathbfcal{OMS}{cmsy}{b}{n}

\titlespacing*{\section}
{0pt}{1ex plus .25ex}{1ex plus 1ex}
\titlespacing*{\subsection}
{0pt}{1ex plus .25ex}{1ex plus 1ex}
\titlespacing*{\subsubsection}
{0pt}{1ex plus .25ex}{1ex plus 1ex}

\titleformat{\section}
 {\fontsize{10}{15}\bfseries }
  {\thesection}
  {1em}
  {}

\titleformat{\subsection}
  {\bfseries }
  {\thesubsection}
  {2em}
  {}

\titleformat{\subsubsection}
  {\itshape}
  {\thesubsubsection}
  {2em}
  {}

\def\beq{\begin{eqnarray}}
\def\eeq{\end{eqnarray}}

\renewcommand{\v}[1]{\ensuremath{\mathbf{#1}}} 
\newcommand{\gv}[1]{\ensuremath{\mbox{\boldmath$ #1 $}}}

\newcommand{\grad}[1]{\gv{\nabla} #1} 
\let\baraccent=\= 
\renewcommand{\=}[1]{\stackrel{#1}{=}} 
\newcommand{\bk}{\boldsymbol{k}} 
\newcommand{\bd}{\textbf{d}} 
\newcommand{\bq}{\textbf{q}} 
\newcommand{\mc}[1]{\mathcal{ #1}} 
\newcommand{\tmc}[1]{\tilde{\mathcal{ #1}}} 
\newcommand{\ra}{\rangle} 

\newcommand{\red}[1]{\textcolor{black}{#1}} 
\newcommand{\bl}[1]{\textcolor{black}{#1}} 

\newenvironment{redp}{\par\color{black}}{\par}

\makeatletter

\titleclass{\subsubsubsection}{straight}[\subsection]

\newcounter{subsubsubsection}[subsubsection]
\renewcommand\thesubsubsubsection{\thesubsubsection.\arabic{subsubsubsection}}

\titleformat{\subsubsubsection}
   {\normalfont\normalsize\itshape \centering}{\thesubsubsubsection}{1em}{}
\titlespacing*{\subsubsubsection}
{0pt}{1ex plus .3ex}{1ex plus .3ex}

\makeatletter
\renewcommand\paragraph{\@startsection{paragraph}{5}{\z@}%
  {3.25ex \@plus1ex \@minus.2ex}%
  {-1em}%
  {\normalfont\normalsize}}
\renewcommand\subparagraph{\@startsection{subparagraph}{6}{\parindent}%
  {3.25ex \@plus1ex \@minus .2ex}%
  {-1em}%
  {\normalfont\normalsize}}
\def\toclevel@subsubsubsection{4}
\def\toclevel@paragraph{5}
\def\toclevel@paragraph{6}
\def\l@subsubsubsection{\@dottedtocline{4}{7em}{4em}}
\def\l@paragraph{\@dottedtocline{5}{10em}{5em}}
\def\l@subparagraph{\@dottedtocline{6}{14em}{6em}}
\makeatother

\begin{document}
\title{Topological skyrmion phases of matter}
\author{Ashley M.\ Cook$^*$}
\affiliation{Department of Physics, University of California, Berkeley,
California, 94720, USA}
\affiliation{Max Planck Institute for Chemical Physics of Solids, N\"othnitzer Strasse 40, 01187 Dresden, Germany}
\affiliation{Max Planck Institute for the Physics of Complex Systems, N\"othnitzer Strasse 38, 01187 Dresden, Germany}
%
%

\begin{abstract}
We introduce topological phases of matter defined by skyrmions in the ground state spin---or pseudospin---expectation value textures in the Brillouin zone, the chiral and helical topological skyrmion phases of matter. These phases are protected by a symmetry present in centrosymmetric superconductors. We consider a tight-binding model for spin-triplet superconductivity in transition metal oxides and find it realizes each of these topological skyrmion phases. The chiral phase is furthermore realized for a parameter set characterizing Sr$_2$RuO$_4$ with spin-triplet superconductivity.  We also find two types of topological phase transitions by which the skyrmion number can change. The second type occurs without the closing of energy gaps in a system described by a quadratic Hamiltonian without breaking the protecting symmetries when atomic spin-orbit coupling is non-negligible and there is a suitable additional degree of freedom. This contradicts the ``flat band'' limit assumption important in use of entanglement spectrum and Wilson loops, and in construction of the ten-fold way classification scheme of topological phases of matter. \red{We furthermore predict two kinds of bulk-boundary correspondence signatures---one for measurements which execute a partial trace over degrees of freedom other than spin, which yields quantized transport signatures---and a second resulting from skyrmions trapping defects with their own non-trivial topology that is discussed in a second work, which yields generalizations of unpaired Majorana zero-modes.}\end{abstract}
\maketitle

\section*{\centerline{\textbf{INTRODUCTION}}}
The search for topologically non-trivial phases of matter---those phases of matter distinguished from one another by changes in appropriate topological invariants rather than symmetry-breaking as in the Ginzburg-Landau paradigm~\cite{chiu2016}---is now a vast and influential topic in condensed matter physics. One of the most important concepts in this research domain is that of the Chern insulator~\cite{haldane1988}, a two-dimensional topologically non-trivial phase of matter central to the construction of many others, from the quantum spin Hall insulator~\cite{kane2005a, kane2005b, bernevig2006, konig766} and the Weyl semimetal~\cite{burkov2011, xubelopolski2015, Lu2015, lv2015}, to topological crystalline phases of matter~\cite{fu2011, hsieh2012}, to topological phases in non-electronic systems~\cite{khanikaev2012, susstrunk2015} and systems out of equilibrium, such as the Floquet topological phases~\cite{lindner2011, cayssol2013, khemani2016}, and dissipative systems, such as the non-Hermitian topological phases~\cite{shen2018, gong2018}.

In this work, we identify and characterize a foundational generalization and counterpart of the Chern insulator, starting from the following concept: the simplest models for Chern insulators are one-electron Hamiltonians with $2 N \times 2 N$ matrix representations and $N=1$. The topology of the Chern insulator is characterized by the first Chern number, computed for the single occupied band when $N = 1$. This topological invariant is also, in this situation, the skyrmion number, the topological charge of the momentum space ground state spin texture (or pseudospin texture depending on the degree of freedom in the model).

For $N>1$, however, this equivalence between the Chern number and the skyrmion number breaks down. For $N>1$, Chern insulators are characterized by a total Chern number $\mc{C}$ of the occupied bands, computed as the sum of the first Chern numbers of individual occupied bands. The skyrmion number $\mc{Q}$, in contrast, is a global topological invariant~\cite{avron1983} and an example of a topological invariant classifying topologically-distinct maps from the Brillouin zone $d$-dimensional torus $T^d$ to the space of ground state spin expectation values $M$, corresponding to non-trivial homotopy group $\pi_d (M) \neq 0$. Thus, while the Chern number and the skyrmion number are locked together for $N=1$, the skyrmion number characterizes topological phases of matter distinct from the Chern insulator for $N>1$ that we introduce in this work. We note that this class of mappings differs from those used for classification, such as in the ten-fold way~\cite{Ryu_2010, schnyder_2008}, which are mappings from the Brillouin zone to the space of projectors.

We present results on two symmetry-protected topological phases of matter. First, we introduce topological phases of matter characterized by a single skyrmion number, or \textit{chiral topological skyrmion insulators} (CTSIs). Second, we present results on two-dimensional topological phases of matter characterized by a pair of skyrmion numbers, or \textit{helical topological skyrmion insulators} (HTSIs). \red{We use this terminology throughout the paper, but here also introduce additional terminology in a revision of the manuscript: we also call the CTSI simply a \textit{Skyrme insulator}, and the HTSI a \textit{helical Skyrme insulator}, as these phases realize a single momentum space spin texture or a pair of such textures similar to counterpart skyrmionic magnetic orders in real-space.}

The CTSI and HTSI can be realized in effectively non-interacting models that may also be interpreted as describing superconductors, by interpreting an off-diagonal block in the matrix representation of the Hamiltonian with particle-hole symmetry as a superconducting gap function (and that Hamiltonian instead as a Bogoliubov de Gennes Hamiltonian). We mean this in the same sense as in Schnyder \emph{et al.}~\cite{schnyder_2008}: BdG symmetry classes DIII, CI, and AIII may be interpreted as superconductors but are termed insulators in this paper, with ``insulator'' referring to the fact that the BCS quasiparticles are fully gapped in the bulk by the mean field pairing gap.

In the context of superconductors, we re-interpret the CTSI and HTSI as the chiral and helical topological skyrmion superconductor phases (CTSS and HTSS, or Skyrme and Antiskyrme superconductors), respectively. Here, we find evidence that both the CTSS and HTSS phases are realized in a model previously used to describe the superconductor Sr$_2$RuO$_4$ in the high-field phase, and that Sr$_2$RuO$_4$ with spin-triplet superconductivity~\cite{Ng_2000, ueno2013} specifically is a candidate for the CTSS phase. We stress, however, that centrosymmetric superconductors can possess the appropriate symmetry to realize a topological skyrmion superconducting phase, either the two-dimensional topological skyrmion phases discussed here or counterpart three-dimensional topological skyrmion phases of matter to be discussed in future work~\cite{liu2020}.

\begin{figure}[t]
\centering
\includegraphics[width=0.45\textwidth]{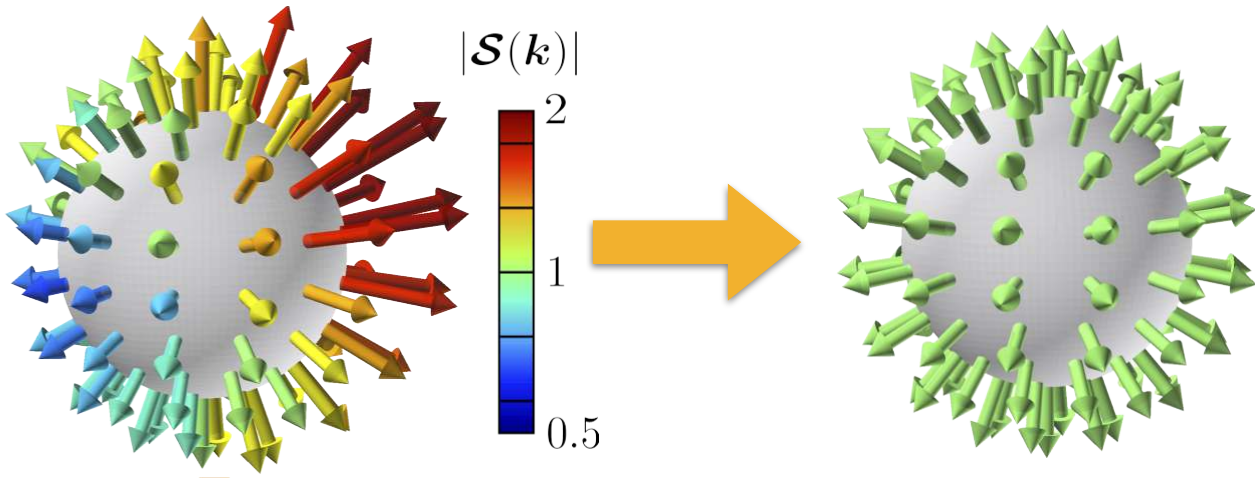}
\caption{\textbf{Mechanism of type II topological phase transition:} Schematic example of unnormalized ground state spin expectation value vector $\boldsymbol{\mc{S}}(\bk)$ texture in the Brillouin zone with skyrmion number of $1$, where the tail of vector $\boldsymbol{\mc{S}}(\bk)$ at momentum $\bk$ in the first Brillouin zone, with orientation given by spherical coordinates $\theta$ and $\phi$, is mapped to point $\left(\theta, \phi \right)$ in spherical coordinates on the two-sphere and shown to the left of the orange arrow for the case of non-negligible atomic spin-orbit coupling. $\boldsymbol{\mc{S}}(\bk)$ as shown to the left of the orange arrow varies in magnitude with $\bk$ in general, due to non-negligible atomic spin-orbit coupling, corresponding to varying color and vector length in this subfigure. When $\boldsymbol{\mc{S}}(\bk)$ is finite everywhere in the Brillouin zone as shown, a skyrmion number is well-defined and can be computed by renormalizing $\boldsymbol{\mc{S}}(\bk)$ by a different value for each $\bk$, as shown to the right of the orange arrow. The skyrmion number can change discretely in the presence of non-negligible atomic spin-orbit coupling by either a type I (with energy gap closing) or a type II topological phase transition (without energy gap closing). The type-II transition occurs because non-neglible atomic spin-orbit coupling in combination with an additional suitable degree of freedom (such as an orbital degree of freedom) allows $\boldsymbol{\mc{S}}(\bk)$ to smoothly pass through zero in magnitude somewhere in the Brillouin zone without a gap closing, causing the skyrmion number computed using the $\hat{\boldsymbol{\mc{S}}}(\bk)$ texture to change discretely from one integer value to another.}
\label{fig0}
\end{figure}

A key feature of these topological phases---a topological phase transition by which the skyrmion number changes without the closing of a gap in the bulk electronic spectrum in an effectively non-interacting system without breaking the symmetries protecting the topological phase---results from the invariant being computed as the topological charge of the ground state spin expectation value in the Brillouin zone when atomic spin-orbit coupling is non-negligible \textit{and} there is a suitable additional degree of freedom (i.e. orbital) vs. negligible atomic spin-orbit coupling or a lack of a suitable orbital degree of freedom, as illustrated in Fig.~\ref{fig0}. In this figure, the tail of the ground state spin expectation value vector at momentum $\bk$ in the first Brillouin zone, $\mc{S}(\bk)$, with orientation $\left(\theta, \phi \right)$ in spherical coordinates is mapped to the point with coordinates $\left(\theta, \phi \right)$ on the two-sphere. Here, the orange arrow represents the process of normalizing $\mc{S}(\bk)$ as required to compute the skyrmion number for each momentum-space spin texture. The spin texture in the Brillouin zone shown in Fig.~\ref{fig0} wraps the two-sphere once, meaning it has a topological charge of $1$.  In a $2N \times 2N$ $(N>1)$ system with non-negligible atomic spin-orbit coupling, the magnitude of the ground state spin expectation value vector $\mc{S}(\bk)$ may vary as a function of $\bk$ if spin is not conserved. This situation is shown in Fig.~\ref{fig0} for the typical case where $\mc{S}(\bk)$ is finite in magnitude everywhere in the Brillouin zone. The topological charge is stable when this vector is finite in magnitude everywhere in the Brillouin zone as the skyrmion number is computed using the normalized, rather than unnormalized, ground state spin expectation value, as illustrated in Fig.~\ref{fig0} to the right of the orange arrow. This vector is finite in magnitude everywhere in the Brillouin zone for large regions of the phase space explored in this work.

When crossing some points in phase space, however, the ground state spin expectation value vector can pass through zero in magnitude smoothly at certain points in the Brillouin zone without the need for closing of an energy gap, permitting a topological phase transition by which the skyrmion number changes discretely from one integer value to another without the closing of any energy gaps in an effectively non-interacting system, without breaking the protecting symmetries of the topological phase. This is the most significant result of the paper as it is the first counterexample to the ``flat band'' limit assumption: given a general band insulator described by a single particle Hamiltonian, a topologically-equivalent ``flat band'' insulator, according to the ``flat band'' limit assumption, can be constructed by setting the energy of all occupied bands to be equal and negative, and all unoccupied bands to be equal and positive through smooth changes of the Hamiltonian while the band gap remains finite~\cite{turner2010}. If a topological phase transition can occur in an effectively non-interacting system without the breaking of the protecting symmetry and without the closing of an energy gap, as we describe here, the ``flat band'' Hamiltonian constructed in this way is no longer guaranteed to be topologically equivalent to the original Hamiltonian in general, and the ``flat band'' limit assumption therefore does not always hold.

%


 The ``flat band'' limit assumption is widespread and foundational in study of topological phases of matter, used, for instance, in the construction of the entanglement spectrum~\cite{fidkowski_2010, turner2010}, Wilson loops~\cite{alexandradinata2014}, and classification schemes such as the ten-fold way~\cite{Ryu_2010, schnyder_2008}. Our work therefore indicates these foundational methods and classification schemes must be reconsidered.

 In this work, we first discuss topological classification and the invariant for the chiral topological skyrmion phases \red{as well as a simple example of a four-band toy model. Four-band toy models are valuable in understanding the non-trivial topology discussed here, but we refer the reader to a second manuscript for in-depth discussion on construction and characterization of four-band toy models of the CTSI and CTSS~\cite{liu2020}.} We then present a general construction of the HTSS in a system with the generalized particle-hole symmetry $\mc{C}'$ and a mirror symmetry. \red{We then show how these symmetries protect the CTSS and HTSS phases in a tight-binding model for superconducting Sr$_2$RuO$_4$.} Phase transitions by which the skyrmion number can change are then discussed, before finishing with discussion of experimental signatures and relevance of these topological phases to topologically-protected quantum computation schemes.

 We note here that, while the chiral topological skyrmion phase may be thought of as a generalization of a Chern insulator, but with the skyrmion number decoupled from the Chern number rather than coupled, the helical topological skyrmion phase may be thought of as a generalization of a quantum spin Hall insulator, which may be constructed as a Chern insulator paired with its time-reversed partner. This makes clear the importance of the helical phase for experimental realization: in the same way the quantum spin Hall insulator can be realized in time-reversal invariant systems without magnetic order, the helical topological skyrmion phase is expected to be realizable in systems without magnetic order. Since the ability to realize a topological phase without magnetic order has greatly facilitated experimental realization in the past (the quantum spin Hall insulator was observed experimentally long before the quantum anomalous Hall insulator~\cite{chang_2013}), the helical topological skyrmion phase is extremely valuable for future study.

  \vspace{6mm}
 \section*{\centerline{\textbf{RESULTS}}}

\section*{
\centerline{Topological classification of the CTSI/CTSS}}

Achieving $\pi_d (M) \neq 0$ for $N>1$ when $M$ corresponds to the space of ground state spin expectation values is a largely unexplored topic. Past work, however, has considered the role of a generalized particle-hole symmetry $\mc{C}'$ in symmetry-protecting the Hopf insulator and a four-dimensional extension of the Hopf insulator in the presence of additional bands corresponding to $\pi_3(Sp(2N)/U(N))$ and $\pi_4(Sp(2N)/U(N))$~\cite{liu2017}. We stress here that, although $\mc{C}'$ symmetry was considered in this past work, the three-dimensional versions of the topological skyrmion phases introduced here were not discussed in that work nor any contradiction to the ``flat band'' limit assumption such as the type II topological phase transition presented here, only the Hopf insulator. Hamiltonians with generalized particle-hole symmetry $\mc{C}'$ acting on a two-dimensional Brillouin zone yield homotopy group $\pi_2(Sp(2N)/U(N)) = \mathbb{Z}$~\cite{liu2017} for mappings from the Brillouin zone to the space of ground state spin expectation values. This implies Hamiltonians with two-dimensional Brillouin zones and $\mc{C}'$ symmetry have $\mathbb{Z}$ topological classification for mapping to this target space. Further details on the topological classification may be found in the Methods, Section I.

We note that $\mc{C}'$ symmetry yields topological classification with dimension $d$ that agrees with the topological classification shown in Table C2 of Ryu \emph{et al.}~\cite{Ryu_2010}, as that table shows topological classification for systems with spatial inversion symmetry and an additional symmetry, including particle-hole symmetry. What was missed previously is that $\mc{C}'$ alone protects an additional topological phase of matter with a topological invariant distinct from those previously considered as we show, which exists only for Hamiltonians with matrix representations of $2N \times 2N$, where $N>1$. Thus, it is possible for the topological skyrmion phases presented here to co-exist with other topological phases of matter, such as those of the ten-fold way. This brings up the interesting possibility of interaction between co-existing topological phases of matter, a topic to be explored in future work.


That the $\mc{C}'$ operator is the product of particle-hole operator $\mc{C}$ and spatial inversion operator $\mc{I}$ suggests a natural place to look for more physically-relevant realizations of these phases is in centrosymmetric superconducting systems, motivating our study here of superconducting Sr\textsubscript{2}RuO\textsubscript{4} in part. Models of Sr\textsubscript{2}RuO\textsubscript{4} with spin-triplet superconductivity in particular possess additional symmetries which we utilize here to find first realization of the helical topological skyrmion phase in addition to the chiral topological skyrmion phase.

 \vspace{6mm}
\section*{
\centerline{Topological invariant and bulk-boundary correspondence}}

As the non-trivial homotopy group has integer topological classification, the topological invariant is a two-dimensional winding number, which may be computed with three-vectors acted upon by ${\mathrm{Sp}(2N) \over \mathrm{U}(N)}$. The question of which quantities may be used to compute the winding number is central to this work. What matters in choosing the appropriate quantity is not whether a quantity is conserved but whether it is acted upon by elements of ${\mathrm{Sp}(2N) \over \mathrm{U}(N)}$ as non-trivial topology of interest here results from the homotopy group $\pi_2 \left( {\mathrm{Sp}(2N) \over \mathrm{U}(N)} \right) = \mathbb{Z}$.

According to the derivation of the homotopy group, it appears, first of all, that we may compute the winding number using projectors onto occupied bands as done in the ten-fold way~\cite{Ryu_2010, schnyder_2008}. However, there is actually another suitable quantity, which is the spin angular momentum. We see that ${\mathrm{Sp}(2N) \over \mathrm{U}(N)}$, where the compact symplectic group can be expressed as the intersection of the complex symplectic group and the special unitary group, $ \mathrm{Sp}(2N, \mathbb{C}) \cap \mathrm{SU}(2N)$, acts on spin given the form of the spin operator matrix representations. Importantly, spin---not total angular momentum (or orbital angular momentum)---is used to compute the winding number even when spin is not conserved. We can see this, by considering that determinants of spin operator matrix representations take the value required by $\mathrm{SU(2N)}$ in general, while neither the determinants for matrix representations of $L$ nor those for matrix representations of $J$ have the correct value in general. This is supported by quantization of the spin skyrmion number over large regions of the phase diagram in the numerics while the topological charge computed instead with total angular momentum $J$ or orbital angular momentum $L$ is everywhere \textit{unquantized} in numerics and therefore topologically trivial.

Skyrmions can therefore form in the texture of the physical spin or pseudospins acted upon by the appropriate quotient, and the corresponding momentum space skyrmionic spin textures observed can be analogous to the spin texture, due to spin-momentum locking, of a Dirac cone in the electronic dispersion at the surface of a topological insulator~\cite{hsieh2008}. More generally, the skyrmion could form in the texture of a pseudospin acted on by $\mathrm{Sp}(2N) / \mathrm{U}(N)$.

A topological invariant can therefore be computed as the winding number (skyrmion number), $\mc{Q}$, of the momentum space ground state spin texture, expressed as an integral over the two-dimensional Brillouin zone,

\begin{equation}
\mc{Q} = {1 \over 4\pi} \int_{BZ} \Omega_{\hat{\mathbfcal{S}}}(\bk) d^2k,
\label{skyrmnum}
\end{equation}

where $\Omega_{\hat{\mathbfcal{S}}}(\bk) = \hat{\mathbfcal{S}}(\bk) \cdot \left( \partial_{k_x} \hat{\mathbfcal{S}}(\bk) \times \partial_{k_y} \hat{\mathbfcal{S}}(\bk) \right)$ is expressed in terms of the normalized ground state spin expectation value $\hat{\mathbfcal{S}}(\bk) = \mathbfcal{S}(\bk) / |\mathbfcal{S}(\bk)|$. We note, additionally, that ${\mathrm{Sp}(2) \over \mathrm{U}(1)} \cong \mathrm{S}^2$, meaning the homotopy group reduces in this case to $\pi_2 \left( \mathrm{S}^2 \right) = \mathbb{Z}$ as expected.

We can also derive this expression by considering projected position operators. This is discussed in Section II of the Methods. Essentially, we may classify the degrees of freedom of a Hamiltonian as spin or pseudospin acted upon by elements of $\mathrm{Sp}(2N) / \mathrm{U}(N)$, $\mathbfcal{S}$, or other degrees of freedom that are not (pseudo)spin, $\bar{\mathbfcal{S}}$. We may then compute the skyrmion connection (curvature) by performing a partial trace on the Berry connection (curvature) of the ground state of the $2N \times 2N$ Bloch Hamiltonian, over $\bar{\mathbfcal{S}}$ degrees of freedom.

Importantly, this reduced connection/curvature yields a bulk-boundary correspondence for non-trivial skyrmion number: in tracing out $\bar{\mathbfcal{S}}$ degrees of freedom, we are left with an effective $2 \times2$ Hamiltonian for the $\mathbfcal{S}$ degree of freedom describing an open system. The boundary is therefore no longer required to respect $\mc{C}'$ symmetry in order to have topologically-robust boundary states as the relevant homotopy group is then $\pi_2(\mathrm{S}^2) = \mathbb{Z}$, which corresponds to lower symmetry. Non-trivial skyrmion number of the ground state spin of the full Hamiltonian therefore yields topologically-protected chiral modes  in the non-equilibrium steady state of the spin subsystem analogous to those of a two-band Chern insulator, when one traces out the other degrees of freedom. Interestingly, we could interpret the spin degree of freedom as a system coupled to an environment consisting of the other degrees of freedom. For large $N$, we could interpret the environment as a reservoir. This suggests the possibility of open systems \textit{appearing} to exhibit non-trivial topology in their non-equilibrium steady state without symmetry-protection, suggesting a possible mechanism for topological order ostensibly similar to that of the fractional quantum Hall effect.

We note that $\mc{C}'$ symmetry restricts the eigenvectors of the Hamiltonian to elements $g = \textrm{diag}\left( U_1(\bk), U_2(\bk) \right)$, where $U_1(\bk)$, $U_2(\bk) \in \textrm{U}(N)$. Additional constraints on $U_2$ relating it to $U_1$ ($U_1^{} U_2^T = I$, where $I$ is the identity matrix, as discussed further in the Methods, Section I) guarantee the existence of a three-vector with which the homotopy invariant may be computed.

One may further confirm the choice by constructing a minimal four band model for a chiral topological skyrmion phase from any two-band Chern insulator with Hamiltonian $\mc{H}(\bk) = d_x(\bk) \sigma_x + d_y(\bk) \sigma_y + d_z(\bk) \sigma_z$, where the Pauli matrices $\{\sigma_i \}$,with $i\in\{0,x,y,z \}$, corresponding to, for instance, a spin degree of freedom. The counterpart minimal four-band model for a chiral topological skyrmion phase is then $\mc{H}_4(\bk) = d_x(\bk) \tau_z\sigma_x + d_y(\bk)\tau_0 \sigma_y + d_z(\bk) \tau_z\sigma_z$, where $\{\tau_i \}$,with $i\in\{0,x,y,z \}$, is a second set of Pauli matrices corresponding to a second degree of freedom beyond spin, such as an orbital degree of freedom. The corresponding spin operators used to compute the skyrmion number are, in this case, $S_i = \mathrm{diag}(\sigma^{}_i, -\sigma_i^*)$, with $i \in \{x,y,z\}$. For half-filling, one may compute a skyrmion number for these four-band models and find that it is decoupled from the total Chern number. For the simple cases outlined above, one finds skyrmion number $Q$ when the total Chern number is $C=-2Q$. These models correspond to superconductors when the normal state Hamiltonian is even in $\bk$, and additional spin-singlet or spin-triplet pairing terms are allowed by $\mc{C}'$ symmetry. This $N=2$ case, corresponding to a four band model, is explored in detail in a separate follow-up work~\cite{liu2020}.

 \vspace{6mm}
\section*{
\centerline{Construction of the HTSS}}

In this work, we study the CTSI phase of matter, characterized by a single non-trivial skyrmion number, $\mc{Q} \neq 0$. We can also construct a second topological phase of matter, the HTSI, which consists of a pair of decoupled CTSIs, one with skyrmion number $\mc{Q}\neq 0$ and the other with skyrmion number $-\mc{Q}$.

One way to effectively construct the HTSI is as follows: we can reinterpret models for the HTSI which possess particle-hole symmetry $\mc{C}$ and spatial inversion symmetry $\mc{I}$, which means the system also has $\mc{C}'$ symmetry, as describing superconductors. If the superconductor also possesses normal state mirror symmetry $\mc{M}$ and superconducting gap function odd under this mirror operation, the superconductor may be divided into two mirror subsectors, which each inherit $\mc{C}$ symmetry~\cite{ueno2013}. We report here that these mirror subsectors of the superconductor also inherit $\mc{C}'$ symmetry under these conditions if the full Hamiltonian possesses $\mc{C}'$ symmetry. This system can then potentially realize the HTSS phase if one mirror subsector has skyrmion number $\mc{Q}$ and the other mirror subsector has skyrmion number $-\mc{Q}$. Details on this construction are discussed in the Methods, Section III.

 \vspace{6mm}
\section*{
\centerline{Tight-binding model realization of the CTSS and HTSS}}

We consider a two-dimensional tight-binding model for Sr$_2$RuO$_4$ with spin-triplet superconductivity on the square lattice~\cite{Ng_2000} also previously used to study topological crystalline superconductivity in this system~\cite{ueno2013}. This model describes conduction electrons of the 4$d$ $t_{2g}$ orbitals of Ru, $d_{yz}$, $d_{xz}$, and $d_{xy}$ in the presence of atomic spin-orbit coupling and spin-triplet pairing. The superconducting state of Sr$_2$RuO$_4$ is described by a Bogoliubov de Gennes Hamiltonian $\mc{H}_{\textrm{BdG}}$ consisting of normal state Hamiltonian $\mc{H} = \mc{H}_{\textrm{kin}}+\mc{H}_{\textrm{soc}} + \mc{H}_{\textrm{B}}$ and gap function $\mc{H}_{\textrm{pair}}$ where
\begin{align}
\mc{H}_{\textrm{kin}} &= \sum_{\bk s}\Psi^{\dagger}_{\bk, s} \mc{H}_{\textrm{kin}_{\hspace{0.5mm} \bk s}} \Psi^{}_{\bk, s} \\
\mc{H}_{\textrm{soc}} &= i \lambda \sum_{\ell m n} \varepsilon_{\ell m n} \sum_{\bk s s'} c^{\dagger}_{\bk s \ell} c^{}_{\bk s' m} \sigma^n_{s s'}\\
\mc{H}_{\textrm{B}} &=  -\mu_B H_z\sum_{k \ell s s'} c^{\dagger}_{\bk s \ell} c^{}_{\bk s' \ell} \sigma^z_{s s'}  \\
\mc{H}_{\textrm{pair}} &= {1 \over 2} \sum_{k \ell s s'} \hat{\Delta}^{\ell}_{s s'}(\bk) c^{\dagger}_{\bk s \ell} c^{\dagger}_{-\bk s' \ell},
\end{align}
with $\Psi^{\dagger}_{\bk, s} = \begin{psmallmatrix}c^{\dagger}_{\bk, s, yz}, & c^{\dagger}_{\bk, s, xz}, & c^{\dagger}_{\bk, s, xy} \\
\end{psmallmatrix}$ and
\begin{equation}
\mc{H}_{\textrm{kin}_{\hspace{0.5mm} \bk s}}= \begin{pmatrix}
\epsilon_{\bk, yz} & g_{\bk} & 0 \\
g_{\bk} & \epsilon_{\bk, xz} & 0 \\
0 & 0 & \epsilon_{\bk, xy} \\
\end{pmatrix}.
\end{equation}

Here, taking lattice spacing constant $a=1$ throughout,
\begin{align}
\epsilon_{\bk, yz} &= -2 t_1 \cos(k_y) - \mu \\
\epsilon_{\bk, xz} &= -2 t_1 \cos(k_x) - \mu \\
\epsilon_{\bk, xy} &= -2 t_2 \left( \cos(k_x) + \cos(k_y) \right) \nonumber \\
&  - 4 t_3 \cos(k_x) \cos(k_y) - \mu'
\end{align}
and $g_{\bk} = -4 t_4 \sin(k_x) \sin(k_y)$. $\hat{\Delta}^{\ell}(\bk) = i \Delta^{\ell} \bd (\bk) \cdot \pmb{\sigma} \sigma_y$, with the $\bd$-vector $\bd(\bk)$ assumed to be $\bd(\bk) = \hat{x} \sin(k_y) - \hat{y} \sin(k_x)$ \textcolor{black}{of the high-field phase, which corresponds to Zeeman field strengths of greater than the critical value of $H_c = 20$ mT} as discussed in Ueno \emph{et al.}~\cite{ueno2013}. Other $\bd$-vectors are possible for the high-field phase and these will be considered in future work.

In this basis, the normal state mirror operator $\mc{M}_{xy}$ taking $z \rightarrow -z$ has matrix representation $\mc{M}_{xy} = \mathrm{diag}\left(-i\sigma_z, -i\sigma_z, i\sigma_z \right)$. $\tmc{M}^-_{xy}$, the appropriate superconductor mirror operator for the high-field phase, may then be constructed as $\tmc{M}^{\pm}_{xy} = \mathrm{diag}\left(\mc{M}^{}_{xy}, \pm \mc{M}_{xy}^* \right)$ as discussed in greater detail in the Methods, Section III.

In the presence of an applied Zeeman field of strength $H_z$ along the $\hat{z}$-axis of Sr$_2$RuO$_4$ (equivalent to the $\hat{c}$-axis in some works), the model for the superconductor possesses particle-hole symmetry $\mc{C}$. We report here that it also possesses the generalized particle-hole symmetry $\mc{C}'$ required for non-trivial homotopy group $\pi_2 (\mathrm{Sp}(2N)/\mathrm{U}(N) ) = \mathbb{Z}$. As well, the gap function is odd under the normal state mirror operation $\mc{M}_{xy}$ taking $z$ to $-z$ in the high-field phase regardless of $\bd$-vector, meaning $\mc{C}'$ is inherited by each counterpart mirror subsector of the superconducting state, corresponding to mirror operator $\tmc{M}^-_{xy}$. In contrast, the low-field phase gap function is even under $\mc{M}_{xy}$, and the $\tmc{M}^+_{xy}$ subsectors do not inherit $\mc{C}'$ symmetry. Thus, the high-field phase of superconducting Sr$_2$RuO$_4$ with spin-triplet pairing satisfies symmetry requirements for the CTSS phase, by having skyrmion number $Q\neq 0$ in one mirror subsector and $Q=0$ in the other mirror subsector, as well as for the HTSS phase, by having $Q\neq0$ in one mirror subsector and $-Q\neq 0$ in the other mirror subsector.


 \vspace{6mm}
\section*{
\centerline{Numerical results}}

We compute the mirror Chern number and the skyrmion number for each $\tmc{M}^-_{xy}$ subsector, or mirror skyrmion number, of the Sr$_2$RuO$_4$ model with spin-triplet superconductivity as a function of spin-orbit coupling $\lambda$ and Zeeman coupling expressed as $\mu_B H_z$, while keeping other model parameters fixed to those used to model Sr$_2$RuO$_4$ specifically in past work~\cite{ueno2013}: $t_1 = t_2 = 0.5$, $t_3 = 0.2$, $t_4 = 0.1$, $\mu = -0.2$, $\mu' = -0.2$, and $\Delta^{\ell} = 0.6$ for each value of $\ell$ ($yz$, $xz$, and $xy$). The mirror Chern number and the mirror skyrmion number are defined in Section IV of the Methods.

To compute mirror skyrmion numbers for the Sr$_2$RuO$_4$ Hamiltonian, we proceed as follows: in the basis in which the superconductor mirror operator $\tmc{M}^-_{xy}$ matrix representation is diagonal, the model for Sr$_2$RuO$_4$ is block-diagonal in the high-field phase. These blocks may be written as $\tmc{H}_1(\bk)$ and $\tmc{H}_2(\bk)$, with corresponding basis vectors $\Psi_1(\bk) = \{c^{\dagger}_{-\bk, xy, \uparrow}, c^{\dagger}_{-\bk, xz, \downarrow}, c^{\dagger}_{-\bk, yz, \downarrow}, c^{}_{\bk, xy, \uparrow}, c^{}_{\bk, xz, \downarrow}, c^{}_{\bk, yz, \downarrow}\}^{t}$, and $\Psi_2(\bk) = \{c^{\dagger}_{-\bk, xy, \downarrow}, c^{\dagger}_{-\bk, xz, \uparrow}, c^{\dagger}_{-\bk, yz, \uparrow}, c^{}_{\bk, xy, \downarrow}, c^{}_{\bk, xz, \uparrow}, c^{}_{\bk, yz, \uparrow}\}^t$, respectively, bases which naturally result from changing to the basis in which $\tmc{M}^-_{xy}$ is diagonal. $\tmc{H}_1(\bk)$ and $\tmc{H}_2(\bk)$ each inherit $\mc{C}'$ symmetry as well as $\mc{C}$ symmetry.

We can construct angular momentum operators for these $\tmc{M}^-_{xy}$ subsectors. First, the orbital angular momentum operator for our bases is $\tilde{\mathbfcal{L}} = \langle \tilde{\mc{L}}_x, \tilde{\mc{L}}_y, \tilde{\mc{L}}_z \rangle$, with
\begin{equation}
\tilde{\mc{L}}_{ i} = \begin{pmatrix} -{\mc{L}}_{ i}^* & 0 \\ 0 & {\mc{L}}_{i} \end{pmatrix}.
\end{equation}
and $i \in \{x, y, z\}$. The corresponding normal state orbital angular momentum operator written in the basis of the $t_{2g}$ orbitals $xy$, $xz$ and $yz$ ~\cite{stamokostas2018} is $\mathbfcal{L} = \langle \mathcal{L}_x,  \mc{L}_y , \mc{L}_z  \rangle$, where
\begin{gather}
\mc{L}_x = \begin{pmatrix}
0 & -i & 0 \\
i & 0 & 0 \\
0 & 0 & 0
\end{pmatrix}, \hspace{5mm}
\mc{L}_y = \begin{pmatrix}
0 & 0 & i \\
0 & 0 & 0 \\
-i & 0 & 0
\end{pmatrix}, \nonumber \\
\mc{L}_z = \begin{pmatrix}
0 & 0 & 0 \\
0 & 0 & -i \\
0 & i & 0
\end{pmatrix}.
\end{gather}
Each $\mc{\tilde{M}}^-_{xy}$ subsector also retains a spin degree of freedom, however, due to the $xy$ being even under $\mc{M}_{xy}$ and the $yz$ and $xz$ orbitals being odd under $\mc{M}_{xy}$. The spin angular momentum operators in the $\hat{x}$-, $\hat{y}$- and $\hat{z}$-directions of the normal state are generalizations of the corresponding spin operators of SU(2), constructed from linear combinations of generators of the Lie group SU(3) ~\cite{georgi1999}. The basis of $\tmc{H}_{1(2)}(\bk)$ corresponds to spin operators $\tilde{\mathbfcal{S}}_{1(2)} = \langle \tilde{\mc{S}}_{1 (2), x}, \tilde{\mc{S}}_{1 (2), y} , \tilde{\mc{S}}_{1 (2), z}  \rangle$, where
\begin{equation}
\tilde{\mc{S}}_{1(2), i} = \begin{pmatrix} -\mc{S}_{1(2), i}^* & 0 \\ 0 & \mc{S}_{1(2), i} \end{pmatrix}.
\end{equation}
Here, the corresponding normal state spin operator is $\mathbfcal{S}_{1(2)} = \langle \mc{S}_{1 (2), x}, \mc{S}_{1 (2), y} , \mc{S}_{1 (2), z}  \rangle$, where (setting fundamental constants to $1$),
\begin{gather}
\mc{S}_{1,x} = {1 \over 2}\begin{pmatrix}
0 & 1 & 1  \\
1 & 0 & 1 \\
1  & 1 & 0
\end{pmatrix}, \hspace{5mm}
\mc{S}_{1,y} = {1 \over 2}\begin{pmatrix}
0 & -i  & -i  \\
i  & 0 & -i \\
i  & i & 0
\end{pmatrix}, \nonumber \\
\mc{S}_{1,z} = {1 \over 2}\begin{pmatrix}
2 & 0 & 0 \\
0 & -1 & 0 \\
0 & 0 & -1
\end{pmatrix}.
\end{gather}
$\tilde{\mathbfcal{S}}_{2}$ can then be expressed in terms of $\tilde{\mathbfcal{S}}_{1}$ using the relations $\mc{S}_{2,x} = \mc{S}_{1,x}$, $\mc{S}_{2,y} = -\mc{S}_{1,y}$, and $\mc{S}_{2,z} = -\mc{S}_{2,z}$, since the raising operator for $\tmc{H}_1(\bk)$ is the lowering operator for $\tmc{H}_2(\bk)$ and vice versa. This derivation of the spin representation from construction of raising/lowering operators within a given mirror subsector is discussed in greater detail in the Methods, Section V. We stress that these are operators for the physical spin and not pseudospin in our case.

\begin{figure}[t]
\centering
\includegraphics[width=0.48\textwidth]{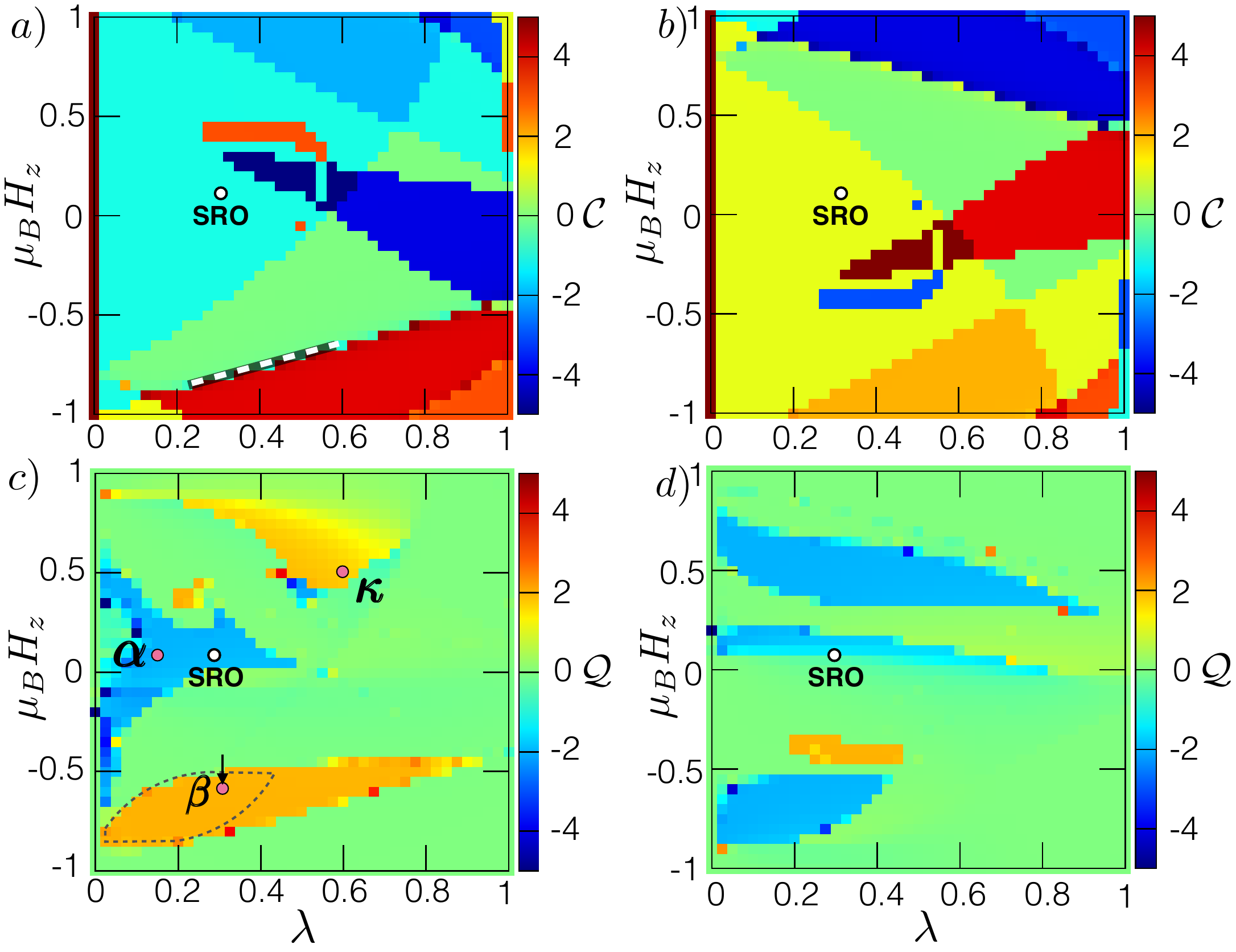}
\caption{\textbf{Topological phase diagrams for mirror subsectors of the Sr\textsubscript{2}RuO\textsubscript{4} tight-binding model:}
Total mirror Chern number $\mc{C}$ computed as a function of spin-orbit coupling $\lambda$ and applied Zeeman field $\mu_B H_z$ with step $\Delta k_i = {\pi \over 150}$ in momentum-space ($i \in \{x, y \}$) for $\tmc{H}_{1}(\bk)$ in subfigure (a) and for $\tmc{H}_{2}(\bk)$ in subpanel (b), respectively.  Mirror skyrmion number $\mc{Q}$ computed as a function of spin-orbit coupling $\lambda$ and applied Zeeman field $\mu_B H_z$ with step $\Delta k_i = {\pi \over 500}$ in momentum-space ($i \in \{x, y \}$) for $\tmc{H}_{1}(\bk)$ in subpanel (c) and for $\tmc{H}_{2}(\bk)$ in subpanel (d), respectively. The parameter set corresponding to Sr$_2$RuO$_4$ is highlighted by the white dot. The three red dots in subfigure (c) show the positions in coordinate space used to compute the momentum space ground state spin textures in Fig.~\ref{fig2}, with Fig.~\ref{fig2} (a), (d), and (g) corresponding to point $\alpha$, Fig.~\ref{fig2} (b), (e), and (h) corresponding to point $\beta$, and Fig.~\ref{fig2} (c), (f), and (i) corresponding to point $\kappa$, respectively. The dashed outline in (c) highlights the region of the phase diagram in which the HTSS phase of matter is realized. The black arrow above point $\beta$ shows the location of the cut shown in Fig.~\ref{fig3}. The black and white dashed line in (a) highlights part of a longer boundary at which the superconducting gap closes. Other model parameters are fixed to those used to model Sr$_2$RuO$_4$ specifically in past work~\cite{ueno2013}: $t_1 = t_2 = 0.5$, $t_3 = 0.2$, $t_4 = 0.1$, $\mu = -0.2$, $\mu' = -0.2$, and $\Delta^{\ell} = 0.6$ for each value of $\ell$ ($yz$, $xz$, and $xy$).}
\label{fig1}
\end{figure}

The total mirror Chern number for $\tmc{H}_1(\bk)$ and the total mirror Chern number for $\tmc{H}_2(\bk)$ as defined in Ueno \emph{et al.}~\cite{ueno2013} are shown in Fig.~\ref{fig1} (a) and (b), respectively, computed as functions of atomic spin-orbit coupling constant $\lambda$ and applied Zeeman field $\mu_B H_z$. These are to be compared with the mirror skyrmion number of $\tmc{H}_1(\bk)$ and the mirror skyrmion number of $\tmc{H}_2(\bk)$, computed using spin operators $\tilde{\mathbfcal{S}}_{1}$ and $\tilde{\mathbfcal{S}}_{2}$, respectively, as functions of spin-orbit coupling constant $\lambda$ and applied Zeeman field $\mu_B H_z$ shown in Fig.~\ref{fig1} (c) and (d). \red{We emphasize here that $\tilde{\mathbfcal{S}}_{1}$ and $\tilde{\mathbfcal{S}}_{2}$ are operators for the physical spin of each mirror subsector, but this does not correspond to block-diagonalizing the spin operators of the full Hamiltonian.} This means spin-ARPES (or a combination of spin-ARPES and CD-ARPES in some cases) is promising for probing these phases, which exhibit topological magnetic order with integer-valued skyrmion numbers, in the same way it has been used to confirm spin-momentum locking of Dirac cones in the surface electronic spectrum of topological insulators~\cite{hsieh2008}. To detect the phases as introduced here, skyrmion numbers must be computed using occupied bands with a particular $\tmc{M}^-_{xy}$ mirror eigenvalue, determined through comparison of experiment with ab initio and tight-binding calculations, using both CD-ARPES and spin-ARPES. (When the CTSS is realized not in a mirror subsector but rather in the full system, spin-ARPES alone can be used without labeling of bands by mirror eigenvalues.) \red{Importantly, quantized transport signatures are also expected to result from non-trival skyrmion number when the transport measurement corresponds to a partial trace over degrees of freedom in the system other than the spin $\mathbfcal{S}$.}

While skyrmion numbers computed using the ground-state spin expectation value may be quantized over regions of the phase diagram as expected from the quotient of the homotopy group, skyrmion numbers computed using the orbital angular momentum operator $\mathbfcal{L}$ or the total angular momentum  $\mathbfcal{J}$ are unquantized and trivial, in contrast. There are some regions where the skyrmion number smoothly transitions from one integer value to another, which will be discussed in greater detail in the next section along with two distinct topological phase transitions by which the skyrmion number changes discretely from one integer value to another.  Noise in the skyrmion number phase diagrams is due to extreme features in the skyrmion number integrand $\Omega_{\bk}$ that can occur over very small regions of momentum-space, necessitating a very fine $\bk$-mesh using known methods.

For a given mirror subsector, $\mc{Q}$ differs in value from $\mc{C}$ and $\mc{Q}$ can be non-zero when $\mc{C}$ is zero and vice versa. $\mc{Q}$ for $\tmc{H}_{1(2)}(\bk)$ can be non-zero when $\mc{Q}$ for $\tmc{H}_{2(1)}(\bk)$ is zero, corresponding to the CTSS phase. Importantly, the CTSS phase occurs even for zero magnetic field, when the system is time-reversal invariant, because $\mc{Q}$ here is computed for mirror subsectors. This CTSS phase of matter occurs for model parameters previously used to model Sr$_2$RuO$_4$ specifically ($\lambda = 0.3$ and $\mu_B H_z = 0.1$ as used in past work~\cite{ueno2013}), indicating Sr$_2$RuO$_4$ is a candidate for the CTSS phase of matter. In the case of Sr$_2$RuO$_4$, the mirror Chern numbers are also non-zero, indicative of two distinct topologically non-trivial phases co-existing. The potential for interaction between these two distinct forms of non-trivial topology is an interesting direction for future work.

In the region of the phase diagram highlighted by a dashed line in Fig.~\ref{fig1} (c), $\mc{Q}$ for $\tmc{H}_{1}(\bk)$ is quantized to $+2$ while $\mc{Q}$ for $\tmc{H}_{2}(\bk)$ is quantized to $-2$. This is the HTSS phase. While time-reversal symmetry is broken in this region of the phase diagram, this phase is analogous to the quantum spin Hall insulator phase constructed as two Chern insulators, one with Chern number $\mc{C} > 0$ and the other with Chern number $-\mc{C}$. Given the potential for the HTSS phase to aid in experimental realization by occurring more generally in the presence of time-reversal symmetry (given it is a generalization of the quantum spin Hall insulator, which occurs in the presence of time-reversal symmetry), this phase is of great significance in its own right.

 \vspace{20mm}
\section*{
\centerline{Nature of the phase transitions}}

\begin{figure}[t]
\centering
\includegraphics[width=0.5\textwidth]{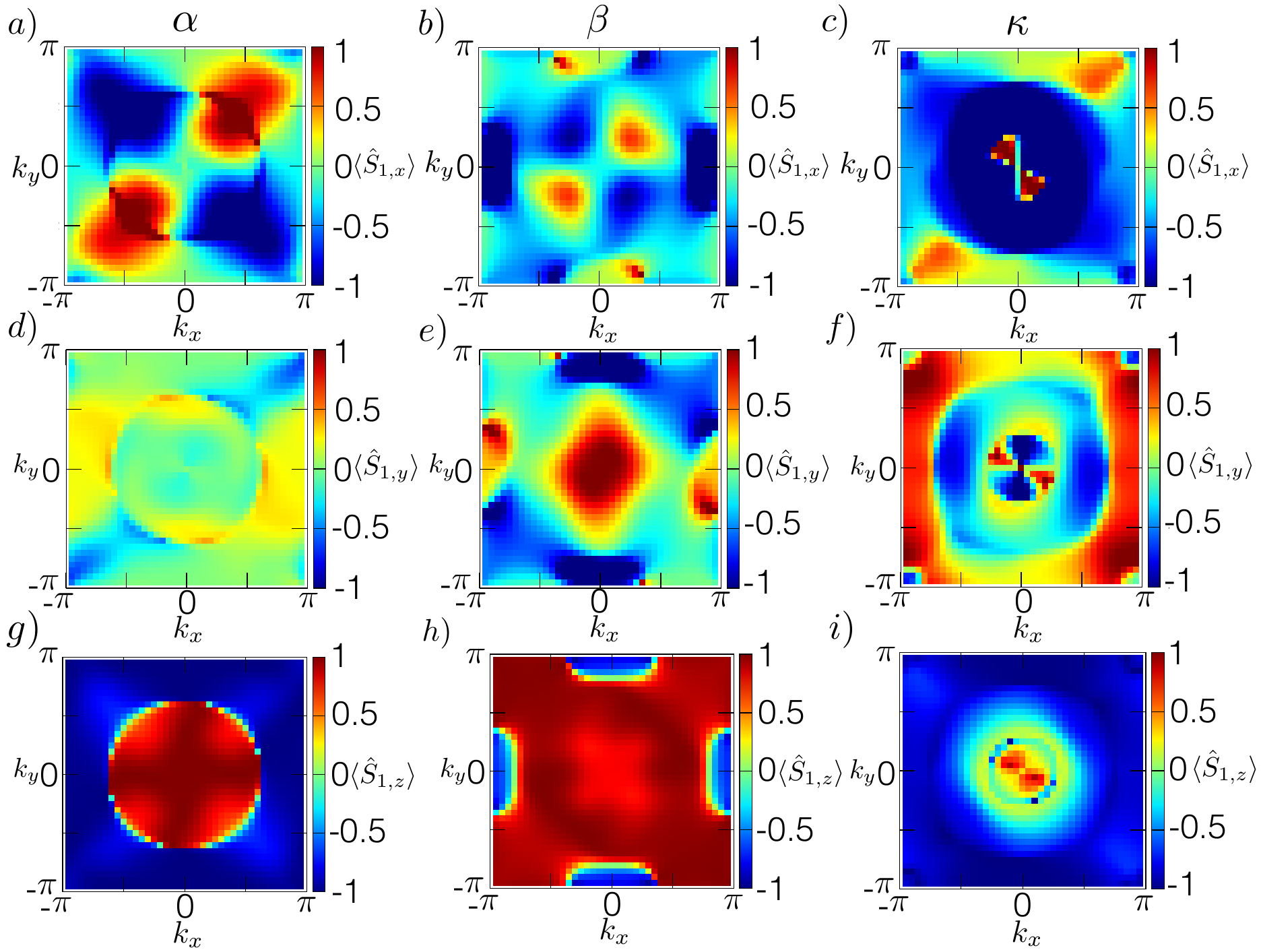}
\caption{\textbf{Examples of skyrmion formation:} Expectation value of the $\hat{x}$-component of spin operator for $\tmc{H}_{1}(\bk)$ , $\langle \hat{S}_{1, x} \rangle$, plotted as a function of $\bk$ at points $\alpha$, $\beta$ and $\kappa$ in phase space, shown in (a), (b), and (c), respectively. Expectation value of the $\hat{y}$-component of spin operator for $\tmc{H}_{1}(\bk)$ , $\langle \hat{S}_{1, y} \rangle$, plotted as a function of $\bk$ at points $\alpha$, $\beta$ and $\kappa$ in phase space, shown in (d), (e), and (f), respectively. Expectation value of the $\hat{z}$-component of spin operator for $\tmc{H}_{1}(\bk)$, $\langle \hat{S}_{1, z} \rangle$, plotted as a function of $\bk$ at points $\alpha$, $\beta$ and $\kappa$ in phase space, shown in (g), (h), and (i), respectively. (a), (d), and (g) were computed for $\lambda = 0.15$ and $\mu_B H_z = 0.1$ and correspond to point $\alpha$ in Fig.~\ref{fig1} (c). (b), (e), and (h) were computed for $\lambda = 0.3$ and $\mu_B H_z = -0.6$ and correspond to point $\beta$ in Fig.~\ref{fig1} (c). (c), (f), and (i) were computed for $\lambda = 0.6$ and $\mu_B H_z = 0.5$ and correspond to point $\kappa$ in Fig.~\ref{fig1} (c). The minimum ground state spin magnitude is zero at $\kappa$, explaining why the topological charge transitions smoothly from $2$ to $0$ in this region of the phase diagram: it is not actually topologically stable in this part of the phase diagram.}
\label{fig2}
\end{figure}

We observe three kinds of phase transitions, two of which correspond to discrete changes in the skyrmion number and one during which the skyrmion number smoothly changes from one integer value to another. To better understand these phase transitions, we first study the momentum space ground state spin textures at points $\alpha$, $\beta$ and $\kappa$ in Fig.~\ref{fig1} (c), shown in Fig.~\ref{fig2}. We find individual skyrmions with $|\mc{Q}|>1$ occur at $\alpha$ and $\beta$ as shown in Fig.~\ref{fig2} (a), (d), and (g) for $\alpha$ and (b), (e), and (h) for $\beta$, respectively. The momentum space ground state spin texture at point $\kappa$, shown in Fig.~\ref{fig2} (c), (f), and (i) instead consists of two features that gradually merge as $\mc{Q}$ transitions from $2$ to $0$ for $\lambda = 0.6$ and increasing $\mu_B H_z$. This non-topological transition occurs because the ground state spin expectation value is zero somewhere in the Brillouin throughout the transition, even though the system formed a spin texture composed of \textit{almost} two fully-formed skyrmions, each with topological charge of $1$.

\begin{figure}[t]
\centering
\includegraphics[width=0.46\textwidth]{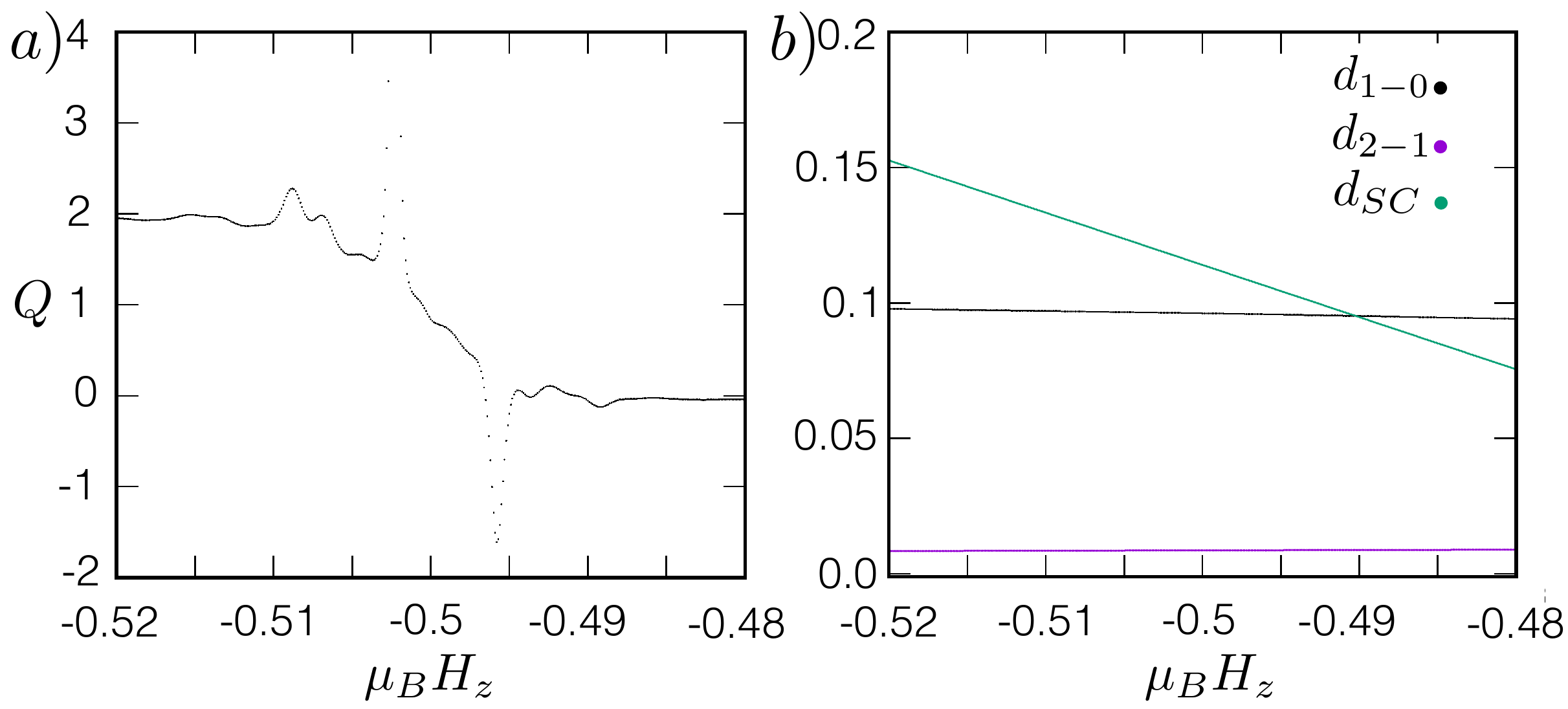}
\caption{\textbf{Type II topological phase transition in greater detail:} (a) Mirror skyrmion number $\mc{Q}$ for $\tmc{H}_{1}(\bk)$ computed as a function of applied Zeeman field $\mu_B H_z$ for fixed spin-orbit coupling value $\lambda = 0.3$ along the cut shown in Fig.~\ref{fig1} (c) with step $\Delta k_i = {\pi \over 300}$, with $i \in \{x, y\}$. (b) The minimum direct gap between the two lowest bands, $d_{1-0}$, the minimum direct gap between the second two lowest bands, $d_{2-1}$, and the minimum direct gap between the two middle bands and minimum direct superconducting gap, $d_{SC}$ of $\tmc{H}_{1}(\bk)$, each computed as a function of applied Zeeman field $\mu_B H_z$ for fixed spin-orbit coupling value $\lambda = 0.3$ along the cut shown in Fig.~\ref{fig1} (c) with step $\Delta k_i = {\pi \over 300}$, with $i \in \{x, y\}$.}
\label{fig3}
\end{figure}

\begin{figure}[t]
\centering
\includegraphics[width=0.48\textwidth]{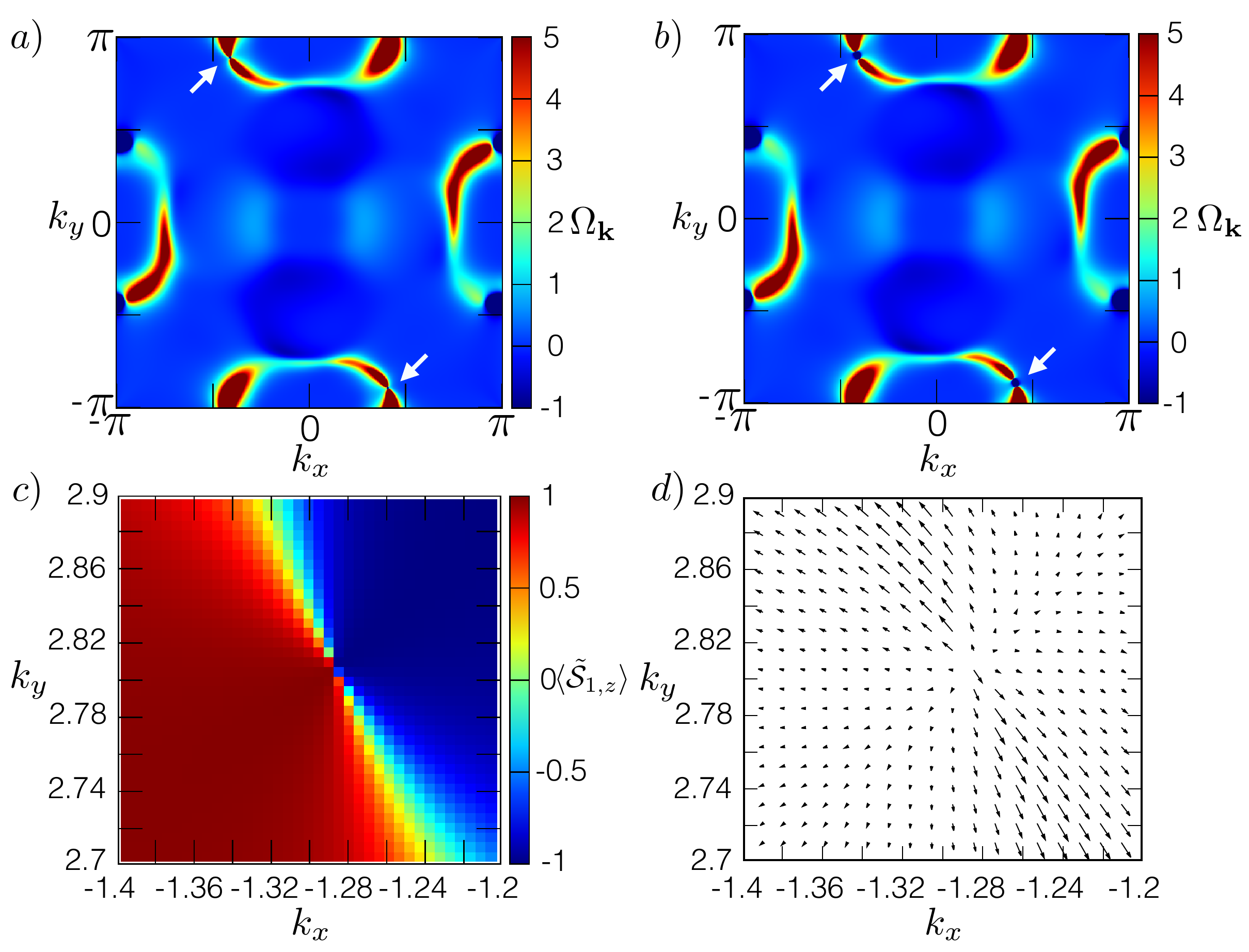}
\caption{\textbf{Numerical evidence of mechanism of type II topological phase transition:} Integrand of the mirror skyrmion number $\mc{Q}$, $\Omega_{\bk}$, as a function of momenta $k_x$ and $k_y$ in the Brillouin zone for spin-orbit coupling constant $\lambda = 0.3$ and applied Zeeman field coupling $\mu_B H_z = -0.5$ shown in (a) and $\mu_B H_z = -0.495$ shown in (b), respectively, two points in the phase diagram along the cut represented by a black arrow in Fig.~\ref{fig1} (c) and further analyzed in Fig.~\ref{fig3}. Two small white arrows in each of (a) and (b) highlight the features (related by symmetry) in $\Omega_{\bk}$ that change during the topological phase transition. For $\mu_B H_z = -0.5$ and $\lambda = 0.3$ corresponding to (a), (c) shows the ground state expectation value of the component of the spin operator in the $\hat{z}$-direction, $\langle \hat{S}_{1, z} \rangle$ and (d)  shows the corresponding expectation values of the components of the spin operator in the $\hat{x}$- and $\hat{y}$-directions, $\langle \hat{S}_{1, x} \rangle$ and $\langle \hat{S}_{1, y} \rangle$, renormalized to prevent overlap of the vectors.}
\label{fig4}
\end{figure}

The two kinds of topological phase transitions are also distinct from one another. In the first kind, type I, the skyrmion number changes with the closing of the superconducting gap, such as along the boundary partially highlighted by the black and white dashed line in Fig.~\ref{fig1} (a), because the ground state spin expectation value computed at the $\bk$-point(s) corresponding to gap-closing(s) is zero and therefore not normalizable, destabilizing the skyrmion number and allowing it to change from one integer value to another. In the second kind, type II, shown in Fig.~\ref{fig3} (a), neither the direct gaps between occupied bands nor the superconducting gap close, as shown in Fig.~\ref{fig3} (b). We note here that the fluctuations in the skyrmion number around the point of the phase transition are due to slow convergence of the skyrmion number noticeable near the phase transition where $\bk$-mesh resolution must be much higher in small regions of the Brillouin zone to better converge to the integer values further away from the phase transition. This slow convergence is also responsible for slight deviation from integer skyrmion number away from the phase transition, which is reduced by increasing $\bk$-mesh resolution. As noted in follow-up work~\cite{liu2020}, much faster convergence for lower-resolution $\bk$-mesh is possible with alternative methods introduced there. The topological nature of this phase transition is clear when Fig.~\ref{fig4} is examined in combination with Fig.~\ref{fig3}. As shown in Fig.~\ref{fig4}, the integrand of the mirror skyrmion number, $\Omega_{\bk}$, during this phase transition changes at two points in the Brillouin zone related by symmetry, highlighted by white arrows in Fig.~\ref{fig4} (a) and (b) for two values of the applied Zeeman field $\mu_B H_z$ between which the change is very noticeable. At the topological phase transition, a red region highlighted by each white arrow forms a neck that is severed into two parts separated from one another by a small, dark blue region. The normalized spin texture in the Brillouin zone near the topological phase transition is shown for the parameter set corresponding to Fig.~\ref{fig4} (a), revealing singularities in the derivative of the normalized ground state spin expectation value develop during the phase transition that destabilize the mirror skyrmion number and permit a discrete change in its value. This singularity in the derivative of the normalized ground state spin expectation value at the topological phase transition corresponds to the unnormalized ground state spin expectation value smoothly going to zero in magnitude at this point in the Brillouin zone (and at the other point in the Brillouin zone related by symmetry) due to spin not being conserved in this system as atomic spin-orbit coupling is non-negligible and there is an additional orbital degree of freedom. Such a process occurs without the breaking of protecting symmetries and may occur without closing of an energy gap due to non-negligible atomic spin-orbit coupling when there is an orbital degree of freedom such that total angular momentum can be exchanged between the spin and orbital sectors.


Since some topological phase transitions can occur in this system without the closing of direct gaps, bulk-boundary correspondence does not necessarily hold for these topologically non-trivial phases. Thus, we do not expect gapless boundary states even for boundaries respecting $\mc{C}'$ symmetry. One possible exception is when a region containing a topological phase may only be reached by type I topological phase transitions, in which the superconducting gap closes.

Such a phase transition in an effectively non-interacting system without breaking the symmetries required for topological protection has important consequences as many results for topological phases assume such phase transitions do not occur. \bl{To our knowledge, the flat-band limit assumption actually first appeared in work by Avron, Seiler, and Simon~\cite{avron1983}, but has been used widely since then.} For instance, connection of the spectrum of the physical Hamiltonian to the entanglement spectrum~\cite{lihui2008, peschel2009, fidkowski2010} and gauge-invariant expressions for Berry connection used for stable numerical computation of Wilson loops~\cite{alexandradinata2014} for a Hamiltonian rely on being able to adiabatically deform that Hamiltonian to a flat band limit that we know is topologically equivalent to the original Hamiltonian. The ten-fold way classification scheme~\cite{Ryu_2010, schnyder_2008} also uses the ``flat band'' limit assumption, because the homotopy groups for classification correspond to mappings from the Brillouin zone to the space of projectors. That is, each class of the ten-fold way corresponds to the set of Hamiltonians that can be adiabatically deformed to a particular flat band Hamiltonian, a projector, without closing the bulk gap. The homotopy group classification is determined from these projector Hamiltonians assuming the ``flat band'' limit assumption holds. The flat-band limit counterpart of the Hamiltonian is only guaranteed to be topologically-equivalent to the original if topological phase transitions only occur with closing of an energy gap. This further suggests that characterization of topological phases with the entanglement spectrum and Wilson loops and use of existing classification schemes reliant on the flat band limit assumption must be done with care: if the system is topologically non-trivial in whole or in part due to topology that can emerge via topological phase transitions without gap closings, the topology characterized by entanglement spectra or Wilson loops may no longer correspond to the topology of the original system and existing classification schemes may no longer apply.

We further note that, while a gap closing is normally expected with a change in skyrmion number in two-band models ($N=1$), this situation is different from $N>1$: there are multiple expressions for the skyrmion number when it is locked to the Chern number, which better reveal why a gap closing is needed to destabilize the calculation in this special case.  The integrand of this skyrmion number for $N>1$ does not take such forms, to our knowledge.

 \vspace{6mm}
\section*{
\centerline{Applications to Sr$_2$RuO$_4$ and other superconductors}}

Recent Knight shift experiments have further called into question the nature of the superconducting pairing in Sr$_2$RuO$_4$~\cite{ pustogow2019, romer2019}, but spin-triplet pairing in Sr$_2$RuO$_4$ remains a very strong possibility~\cite{mackenzie2017}. While the parameter set used here for Sr$_2$RuO$_4$ in the high-field phase is established by previous works, the topology realized for other $\bd$ vectors and further characterization of the parameter values relevant to Sr$_2$RuO$_4$ should be explored.

The HTSS phase, while occurring for a value of the spin-orbit coupling constant $\lambda$ relevant to Sr$_2$RuO$_4$, occurs for strong Zeeman fields at which the parameter $\Delta$ of the superconducting gap function is expected to be zero in experiments. This work instead primarily serves as the introduction of this phase and its realization in a physically-relevant model, although the phase does persist to smaller superconducting pairing strength $\Delta$.

Given that the $\mc{C}'$ operator is the product of particle-hole operator $\mc{C}$ and spatial inversion operator $\mc{I}$, $\mc{C}'$ symmetry is widespread. Thus, this work serves as motivation for a broad search for other candidates for the topological skyrmion phases of matter. This work also serves as further motivation and guidance in the search for counterpart higher-dimensional topological skyrmion phases of matter, which will be characterized in part by the invariant presented here. The three-dimensional topological skyrmion phases, for instance, should be characterized by three weak invariants, the invariant discussed here, and one strong invariant, the topological charge of a three-dimensional skyrmion in the momentum-space spin texture.

Follow-up work on three-dimensional topological skyrmion phases~\cite{liu2020} shows non-trivial skyrmion invariant can yield bulk-boundary correspondence, even when the standard topological invariant computed as the winding of a projector onto occupied states is trivial: the 3D topological skyrmion phase can protect unpaired Majorana bound-states~\cite{liu2020}. Experiments measuring tunneling conductance or transport signatures may therefore be relevant in detecting topological skyrmion phases, as well as spin-resolved ARPES---possibly in combination with CD-ARPES---in certain cases, which can reveal signatures of the momentum-space skyrmions defining these topological phases (in some cases, soft x-ray ARPES techniques in particular may be useful for probing $\boldsymbol{k}$-space spin textures in bulk~\cite{Fedchenko_2019}). Related to this, we comment briefly on the expected emergence of gapless boundary modes within the bulk gap during a type-II topological phase transition from a trivial topological skyrmion phase, without topologically-robust boundary modes in the bulk gap, to a non-trivial one with topologically-robust boundary modes in the bulk gap. The minimum direct bulk gap is not expected to close. The boundary modes are instead expected to descend into the bulk gap and be pinned to zero energy at the point of the type-II topological phase transition and within the non-trivial topological skyrmion phase.

  \vspace{6mm}
 \section*{\centerline{\textbf{DISCUSSION AND CONCLUSION}}}

We have introduced symmetry-protected topological phases of matter in one-electron Hamiltonians with $2 N \times 2 N$ matrix representations and $N>1$ characterized by ground state spin expectation value textures in the Brillouin zone forming skyrmions, the chiral and helical topological skyrmion phases of matter. These topological phases of matter are protected by a generalized particle-hole symmetry $\mc{C}'$ present in some centrosymmetric superconductors and corresponding to non-trivial homotopy group $\pi_2(Sp(2N)/U(N)) = \mathbb{Z}$ for mappings from the Brillouin zone to the space of ground state spin expectation values. They are characterized by a global topological invariant, the skyrmion number, $Q$, computed as the topological charge of the momentum-space ground state spin expectation value texture. While for Hamiltonians with $N=1$, this skyrmion number is locked in value to the Chern number, the skyrmion number decouples from the Chern number for $N>1$ and characterizes phases of matter distinct from the Chern insulator. Most notably, strikingly different phase transitions by which the skyrmion number can change are possible for $N>1$. In systems with non-negligible spin-orbit coupling, for instance, a type II topological phase transition is possible by which the skyrmion number changes discretely from one integer value to another as the ground state spin expectation value passes smoothly through zero in magnitude at a small set of points in phase space, corresponding to a topological phase transition in an effectively non-interacting system without the closing of energy gaps and without breaking of protecting symmetries. This serves as the first counterexample to the ``flat band'' limit assumption foundational to study of topological phases of matter, used, for instance, in the construction of the entanglement spectrum~\cite{fidkowski_2010, turner2010}, Wilson loops~\cite{alexandradinata2014}, and the ten-fold way classification scheme~\cite{Ryu_2010, schnyder_2008}. Our work therefore indicates more general methods sensitive to type II topological phase transitions should be sought in efforts to characterize topological phases of matter.

This type II topological phase transition which occurs without the closing of energy gaps in an effectively non-interacting system without breaking protecting symmetries can be understood as follows: in a system with non-negligible spin-orbit coupling, the unnormalized ground state spin expectation value may change in magnitude as shown in Fig.~\ref{fig0} b) since spin is not conserved. Since the skyrmion number is computed with the normalized ground state spin expectation value vector, the skyrmion number is well-defined so long as the magnitude of the vector is finite everywhere in the Brillouin zone. The minimum spin magnitude over the Brillouin zone can smoothly pass through zero at a small set of points in phase space without closing of energy gaps, however, in which case the skyrmion number can change discretely from one integer value to another due to singularities in the derivative of the normalized ground state spin expectation value at these points in $\bk$-space, corresponding to a topological phase transition.

\begin{redp}
That this type-II topological phase transition indicates bulk-boundary correspondence is not necessarily satisfied for the topological skyrmion phases of matter\red{---\textit{if measurement does not include a partial trace over degrees of freedom other than spin}---}is in line with past understanding that topological phases of matter are not required to satisfy bulk-boundary (sometimes also referred to as bulk-edge) correspondence: a very well-known example is the $\mathbb{Z}_2$ quantum spin liquid, a topological phase of matter for which bulk-boundary correspondence does not hold. The $\mathbb{Z}_2$ spin liquid phase is characterized by a topological invariant that can be changed while keeping the bulk gap finite~\cite{cong2017, levin2013, wang2015}, corresponding to the absence of protected gapless edge modes when the system is topologically non-trivial~\cite{hasan2010}.

\red{In addition, when measurements on the system perform a partial trace over all degrees of freedom except for spin, we predict quantized transport signatures in the non-equilibrium steady state of the resultant open system, associated with non-trivial skyrmion numbers characterizing the CTSI and HTSI. For superconductors, we specifically predict quantized thermal Hall conductivity when measurements execute the appropriate partial trace, corresponding to discarding knowledge about degrees of freedom other than spin.}

We note that the skyrmion number being integer-quantized in only part of the phase diagram \textit{does not prevent it} from serving as a topological invariant in the regions where it is integer quantized. In the regions of the phase diagram where the skyrmion number is everywhere quantized to a particular non-trivial integer value, and must change discretely to another integer value when crossing the boundary enclosing this region, the skyrmion number is a physical quantity unaffected by sufficiently small perturbations, and therefore a topological invariant characterizing a topological phase of matter~\cite{hasan2010}. A simple example of this is the intrinsic Hall conductivity, which is sometimes quantized in integer multiples of $e^2/h$ corresponding, for instance, to the integer quantum Hall effect, but can also be unquantized, such as in a metal~\cite{Haldane2004}. This lack of quantization in some situations does not prevent the intrinsic Hall conductivity from serving as a topological invariant in others.
\end{redp}

We furthermore find both phases are realized in an established tight-binding model describing spin-triplet superconductivity in Sr$_2$RuO$_4$, in this context denoted by chiral topological skyrmion superconductor (CTSS) and helical topological skyrmion superconductor (HTSS), respectively. The CTSS phase is furthermore realized for a parameter set previously established for Sr$_2$RuO$_4$ in the high-field phase, indicating Sr$_2$RuO$_4$ may harbor topological skyrmion phases. Three different kinds of phase transitions are observed by which the skyrmion number changes, two of which are topological, and one which is smooth, differentiated by the structure of the momentum space ground state spin texture in the non-trivial phase. Topological phase transitions can also occur with or without closing of direct gaps in this system modeled by a quadratic Hamiltonian, with important implications in study of topological physics very generally.

\red{As well, these topological phases are able to trap generalizations of unpaired Majorana zero-modes, which we call cross zero-modes, as we discuss in follow-up work that introduces the three-dimensional topological skyrmion phases~\cite{liu2020}. We comment on this additional work here only briefly: the three-dimensional skyrmions are able to trap defects in the Brillouin zone, and such defects can themselves realize topologically non-trivial phases as discussed in Teo and Kane~\cite{teoprl2010, teo2010}, which can themselves exhibit a bulk-boundary correspondence. Such a bulk-boundary correspondence can realize what we call ``cross zero-modes'', which can be thought of as single states consisting of an unpaired Majorana zero-mode and its $\mc{C}'$ partner~\cite{teo2010}. In particular, a four-band Bloch Hamiltonian for a three-dimensional chiral skyrmion phase may be constructed from a Hopf insulator Bloch Hamiltonian, its $\mc{C}'$ partner, and a spin-triplet pairing function, as we show in this second work. The cross zero-modes are then realized using the same boundary conditions under which unpaired Majorana zero-modes may be realized for the Hopf insulator~\cite{yanzhongbo2017}}.



\red{In conclusion, given the tremendous interest in skyrmions for fundamental understanding and applications in many very high-impact works~\cite{fert2017, roessler2006, muhlbauer2009, romming2013}, these results serve as a valuable starting point in greater understanding of a very large area of condensed matter research on skyrmion physics and its applications. Experimental detection of these skyrmions in the Brillouin zone via spin-ARPES or a combination of CD-ARPES and spin-ARPES for skyrmions in mirror subsectors~\cite{fedchenko2019}---as well as transport measurements to detect the bulk-boundary correspondence associated with the skyrmions, in particular thermal transport measurements in superconductors---is therefore of great significance.} The possible realization of this non-trivial topology in Sr$_2$RuO$_4$, an intensely-studied superconductor, is itself also valuable, motivating review of past results on the material, as well as investigation of other superconductors as hosts to this physics given that the $\mc{C}'$ symmetry can be respected by centrosymmetric superconductors.

\vspace{6mm}
\section*{
\centerline{\textbf{ACKNOWLEDGEMENTS}}}

We thank S.~Raghu, S.~A.~Kivelson, A.~Mackenzie, R.~Moessner, B.~Doucot, Y.~B. Kim, J. Motruk, F. de Juan, M. Sigrist, T. Schuster, M. Zaletel, D. Podolsky, S. Gazit, J. Katoch, R. Mong, C. Timm, and E. Altman for useful discussions. We especially wish to acknowledge J.~E. Moore for many helpful discussions. A.~M.~C. also wishes to thank the Aspen Center for Physics, which is supported by National Science Foundation grant PHY-1066293, and the Kavli Institute for Theoretical Physics, which is supported by the National Science Foundation under Grant No. NSF PHY-1125915, for hosting during some stages of this work. A.~M.~C. was supported by the NSERC PDF.

\vspace{6mm}
\section*{
\centerline{\textbf{METHODS}}}

\textbf{I. Topological classification of the CTSI.}

We first review and extend work by Xu and co-authors~\cite{liu2017} to show a generalized particle-hole symmetry $\mc{C}'$ is desired for realization of the homotopy group $\pi_2(Sp(2N)/U(N))$. From work by Bott~\cite{bott1959}, we have the following relation between homotopy groups, where Sp is the compact symplectic group and U is the unitary group,
\begin{equation}
\pi_k ( \textrm{Sp} / \textrm{U} ) = \pi_{k+1} (\textrm{Sp}).
\end{equation}

Xu and co-authors~\cite{liu2017} use this relation in combination with work on the homotopy groups of symplectic groups~\cite{mimura1963} to determine the topological classification of the set of Hamiltonians they present, with homotopy group $\pi_3(Sp(2N)/U(N))$, to be $\mathbb{Z}_2$ for $N>1$. They also briefly consider $\pi_4(Sp(2N)/U(N))$.

To understand whether the generalized particle-hole symmetry $\mc{C}'$ is desired in the case of a two-dimensional Brillouin zone for Hamiltonians to be topologically-classified by homotopy group $\pi_2 (\mathrm{Sp}(2N)/\mathrm{U}(N) )$, we consider a Hamiltonian $H(\bk)$ possessing $\mc{C}'$ symmetry. $H(\bk)$ therefore satisfies

\begin{equation}
\mc{J}^{-1} H(\bk) \mc{J} = -H^*(\bk)
\end{equation}
for each $\bk$, with $\mc{J} =  \begin{psmallmatrix} 0 & I_{N\times N} \\ -I_{N\times N} & 0 \end{psmallmatrix}$, where $I_{N\times N}$ is the $N\times N$ identity matrix. This constraint on the Hamiltonian for each $\bk$ restricts the set of Hamiltonians permitted to the symplectic group, Sp, and also forces the number of bands of the Hamiltonian to be even.

A generic element that does not move $I_{N\times N}$ is $g = \textrm{diag}\left( U_1(\bk), U_2(\bk) \right)$, where $U_1(\bk)$, $U_2(\bk) \in \textrm{U}(N)$. For $g$ to be in $\textrm{Sp}(2N)$, it must satisfy

\begin{equation}
\begin{pmatrix}
U_1 & 0 \\
0 & U_2 \\
\end{pmatrix}
\mc{J}
\begin{pmatrix}
U_1 & 0 \\
0 & U_2 \\
\end{pmatrix}^T
= \mc{J}.
\end{equation}
For this expression to hold, $U_1^{} U_2^T = I$, where $I$ is the identity matrix. This corresponds to the configuration space of the Hamiltonian being $\mathrm{Sp}(2N) / \mathrm{U}(N)$. Xu's discussion~\cite{liu2017} holds whether we consider a three-dimensional Brillouin zone or a two-dimensional one. Thus, we expect Hamiltonians with two-dimensional Brillouin zones and $\mc{C}'$ symmetry to have $\mathbb{Z}$ topological classification characterized by a global homotopy invariant.
\vspace{2mm}

\redp{\textbf{II. Derivation of the spin skyrmion number invariant using projected position operators and the observable-enriched partial trace.}

This discussion extends Parameswaran, Roy and Sondhi~\cite{parameswaran_roy_sondhi}. There, they explore the restriction of kinematics to a single Chern band. We instead restrict to projections of the density operator onto the ground state manifold. Then, differentiating between the spin degree of freedom, $\mathbfcal{S}$, and all other degrees of freedom of the Hamiltonian, $\bar{\mathbfcal{S}}$, we perform a partial trace over degrees of freedom $\bar{\mathbfcal{S}}$.

An $N$-band, single-particle Hamiltonian may be written as
\begin{align}
    \mc{H} = \sum_{\bk,a,b} c^{\dagger}_{\bk, a}\mc{H}_{ab}(\bk)c^{}_{\bk, b},
    \end{align}
where $a,b = 1, 2, ..., N$ are spin and other indices, and $\bk$ is the crystal momentum restricted to the first Brillouin zone (BZ). The solution to the $N \times N$ eigenvalue problem for band $\alpha$ is $\sum_{b} \mc{H}^{}_{ab}(\bk) u^{\alpha}_{b}(\bk) = E_{\alpha}(\bk) u^{\alpha}_a(\bk)$. We take these  eigenvectors to be normalized, $\sum_a |u_a^{\alpha}(\bk)|^2 = 1$. We then define a projector onto the ground state manifold, $\mc{P}_{GS}$, as $\mc{P}_{GS} = \sum_{\alpha,\beta}|\bk, \alpha \rangle \langle \bk, \beta |$, where

\begin{align}
|\bk, \alpha\rangle = \gamma^{\dagger}_{\bk,\alpha} |0 \rangle \equiv \sum_{a} u^{\alpha}_{a} (\bk) c^{\dagger}_{\bk, a} |0 \rangle
\end{align}

The density operator for the full system is $\rho_{\bq} = e^{i\bq \cdot \boldsymbol{r}}$. The density operator projected onto the ground state manifold is then

\begin{align}
\bar{\rho}_{\bq} &= \mc{P}_{GS} \rho(\bq) \mc{P}_{GS} \nonumber \\  &= \sum_{\bk}\sum_{\alpha, \beta \in GS} \left(\sum_{a,b}^N u^{{\alpha}*}_{a}\left(\bk + {\bq \over 2} \right)  u^{\beta}_{b}\left(\bk - {\bq \over 2} \right)\right)  \\ \nonumber
&\times \gamma^{\dagger}_{\bk+{\bq \over 2}, \alpha} \gamma^{}_{\bk-{\bq \over 2}, \beta}
\end{align}

We now consider the reduced density matrix $\bar{\rho}_{\bq,\mathbfcal{S}}$, where
\begin{align}
\bar{\rho}_{\bq,\mathbfcal{S}} = \mathrm{\tilde{T}r}_{\bar{\mathbfcal{S}}}\bar{\rho}_{\bq}.
\end{align}

 Here, $\mathrm{\tilde{T}r}_{\bar{\mathbfcal{S}}}$ is an \textit{observable-enriched} partial trace introduced in a follow-up work~\cite{winter2023}, over all degrees of freedom that are not spin, $\bar{\mathbfcal{S}}$. To briefly summarize the key result there, the observable-enriched partial trace is the operation performed on $\bar{\rho}_{\bq}$ such that $\bar{\rho}_{\bq,\mathbfcal{S}}$ satisfies the following equality between expectation value for ground state spin in the full system in terms of spin operator matrix representation $\hat{\mathbfcal{S}}$, $\mathrm{Tr}\left(\bar{\rho}_{\bq} \hat{\mathbfcal{S}}\right)$, and expectation value for ground state spin in the system after tracing out all degrees of freedom except spin, $\mathrm{Tr}\left(\bar{\rho}_{\bq,\mathbfcal{S}} \hat{\boldsymbol{\sigma}}\right)$, which uses spin operator matrix representation $\hat{\boldsymbol{\sigma}}$,
 \begin{align}
\mathrm{Tr}\left(\bar{\rho}_{\bq} \hat{\mathbfcal{S}}\right) = \mathrm{Tr} \left( \bar{\rho}_{\bq,\mathbfcal{S}} \hat{\boldsymbol{\sigma}} \right).
 \end{align}
For spin-$1/2$, $\hat{\boldsymbol{\sigma}}$ is the vector of Pauli matrices.

 As in Parameswaran~\emph{et al.}, we consider long wavelengths $\bq \cdot \boldsymbol{a} \ll 1$, in which case we expand

\begin{align}
\bar{\rho}_{\bq,\mathbfcal{S}}&= \mathrm{Tr}_{\bar{\mathbfcal{S}}} \left[\sum_{\alpha, \beta \in GS}\sum_{a,b}^N u^{\alpha*}_a \left(\bk + {\bq \over 2} \right) u^{\beta}_b \left(\bk - {\bq \over 2} \right) \right]  \nonumber \\
&\approx 1 - i \bq \cdot \mathrm{Tr}_{\bar{\mathbfcal{S}}} \left[\sum_{\alpha, \beta \in GS} \sum_{a,b}^N u^{\alpha*}_a \left(\bk \right) { \grad_{\bk} \over i} u^{\beta}_b \left(\bk  \right) \right] \nonumber \\
&\approx e^{i \int_{\bk - \bq/2}^{\bk+\bq/2} d\bk ' \cdot \mathrm{Tr}_{\bar{\mathbfcal{S}}}\left[\sum_{\alpha,\beta \in GS}\mc{A}_{\alpha, \beta}(\bk ') \right]}.
\end{align}

We may then consider how the reduced density matrix operator $\bar{\rho}_{\bq,\mathbfcal{S}}$ acts on a state $|\bk, \mathbfcal{S}\rangle$ of the system after the partial trace has been performed, which is

\begin{align}
\bar{\rho}_{\bq,\mathbfcal{S}} |\bk, \mathbfcal{S}\rangle \approx e^{i \int_{\bk - \bq/2}^{\bk+\bq/2} d\bk ' \cdot \mc{A}^{\mathbfcal{S}}(\bk ')} |\bk +\bq,\mathbfcal{S} \rangle.
\end{align}
where here we consider a spin-$1/2$ degree of freedom, so $|\bk, \mathbfcal{S}\rangle$ is an eigenstate of a two-band Bloch Hamiltonian which characterizes the ground state spin of the $N\times N$ problem. This corresponds to $\bar{\rho}_{\bq,\mathbfcal{S}}$ implementing parallel transport for the ground state spin, described by the reduced connection $\mc{A}^{\mathbfcal{S}}(\bk )$.

Although the dynamics of the full Bloch Hamiltonian describing an isolated system are unitary, the dynamics of a subsystem corresponding to a partial trace over some degrees of freedom are not strictly unitary in general. We could interpret the full Bloch Hamiltonian as consisting of a system containing just the spin degree of freedom, and an environment, corresponding to the remaining degrees of freedom. Unlike in the typical scenario for open systems, however, our environment is similar in scale to our system. Thus, in the cases we are considering, of Bloch Hamiltonians with $2N \times 2N$ matrix representation and $N$ relatively small, it is not appropriate to consider a Lindblad master equation~\cite{lieu2020}, although this would be more relevant for large $N$ corresponding to a reservoir with far more degrees of freedom. In our case, dynamics of the reduced Berry connection can be determined at each step in the evolution of the full Berry connection by performing the partial trace.

For a spin-$1/2$ degree of freedom, we may therefore write a two-band Bloch Hamiltonian that could be used to compute this reduced connection $\mc{A}^{\mathbfcal{S}}(\bk )$ in terms of the ground state spin expectation value vector of the full system, $\langle \mathbfcal{S}_{GS}(\bk) \rangle$, as

\begin{align}
    \mc{H}_{\mathbfcal{S}}(\bk) = \langle \mathbfcal{S}_{GS}(\bk) \rangle \cdot \boldsymbol{\sigma},
\end{align}

where $\boldsymbol{\sigma}$ is the vector of Pauli matrices. The skyrmion number of the spin for such a two-band Hamiltonian may also be expressed in terms of a reduced curvature  $\Omega^{\mathbfcal{S}}(\bk ) = \grad \times \mc{A}^{\mathbfcal{S}}(\bk )$. For our purposes, it is sufficient to integrate a single component of $\Omega^{\mathbfcal{S}}(\bk )$ relevant to the $k_x$-$k_y$ plane to compute our skyrmion number,

\begin{align}
\mc{Q} &= {1\over 4 \pi} \int_{BZ} {d\bk \over |\mc{\mathbfcal{S}}_{GS}(\bk)|^3} \left[ \langle \mc{\mathbfcal{S}}_{GS} (\bk) \rangle \cdot  \left( \partial_{k_x} \langle \mc{\mathbfcal{S}}_{GS} (\bk)  \rangle \right. \right. \nonumber \\
&\left. \left. \times \partial_{k_y} \langle \mc{\mathbfcal{S}}_{GS} (\bk)  \rangle \right) \right].
\end{align}

As a result of the $\mc{C}'$ symmetry of the full Bloch Hamiltonian, this skyrmion number is quantized when the spin is finite in magnitude everywhere in the Brillouin zone, and the reduced Berry phase for the spin subsystem is therefore also quantized despite coupling to the environment. The response signatures of the spin subsystem expressed in terms of the skyrmion number are therefore as robust as in the case of purely unitary evolution. Measurements corresponding to a partial trace furthermore effectively reduce the homotopy group to $\pi_2(\mathrm{S}^2) = \mathbb{Z}$, so a non-trivial skyrmion number $\mc{Q}$ for the spin texture in the Brillouin zone can then yield $|\mc{Q}|$ topologically-protected chiral modes in the non-equilibrium steady state of the spin subsystem. In effect, non-trivial skyrmion number corresponds to the system realizing a Chern insulator phase characterized by a two-band Bloch Hamiltonian \textit{specifically for the spin degree of freedom, upon tracing out other degrees of freedom}. Skyrmion number $\mc{Q} \neq 0$ therefore corresponds to $\mc{Q}$ chiral modes on each edge in the spin subsystem with open boundary conditions. The system before partial trace is constrained by the spin degree of freedom realizing this topological phase and by its bulk-boundary correspondence. That is, the spin texture at the edge in the full system must yield the chiral modes of the spin subsystem after the observable-enriched partial trace. In a two-band Chern insulator, chiral modes exhibit very strict spin-momentum-locking, and a skyrmion phase leads to spin-momentum-locking persisting in systems with greater than two bands. Further details are included in our follow-up work on the observable-enriched partial trace~\cite{winter2023}. More generally, these results apply to response functions of the spin subsystem, which can be expressed in terms of these reduced geometric quantities resulting from performing the partial trace.

For large $N$ at which the Lindbladian perspective becomes more suitable, this would correspond to additional restrictions on the dissipators due to symmetries of the system plus environment. We will explore this in future work.

\textbf{III. Construction of the HTSS.}

First, we consider a schematic model for a two-dimensional superconductor in the $x-y$-plane with Bogoliubov-de Gennes (BdG) Hamiltonian as in Ueno~\emph{et al.} ~\cite{ueno2013}, expressed as
\begin{equation}
\mc{H} = \sum_{\bk} \Psi^{\dagger}_{\bk} \mc{H}(\bk) \Psi_{\bk}/2,
\end{equation}
where
\begin{equation}
\mc{H}(\bk) = \begin{pmatrix}
\mc{E}(\bk) & \Delta(\bk) \\
\Delta^{\dagger}(\bk) & -\mc{E}^{T}(-\bk) \\
\end{pmatrix}
\end{equation}
and $\Psi_{\bk} = \left(c^{}_{\bk s \ell}, c^{\dagger}_{-\bk s \ell} \right)^t$. Here, $c^{}_{\bk s \ell}$ is the annihilation operator of electrons with momentum $\bk = \left( k_x, k_y \right)$, spin angular momentum degree of freedom $s \in \{ \uparrow, \downarrow \}$, and orbital angular momentum degree of freedom $\ell \in \{ 1, ..., N \}$. $\mc{E}(\bk)$ is the Hamiltonian of the normal state, and $\Delta(\bk)$ is the gap function of the superconductor.

We consider a normal state with a mirror symmetry, corresponding to $\mc{E}(\bk)$ satisfying the relation $\mc{M}_{xy}^{} \mc{E}(\bk) \mc{M}_{xy}^{\dagger} = \mc{E}(\bk)$, where $\mc{M}_{xy}$ is the matrix representation of the mirror operation taking $z \rightarrow -z$.

The superconducting state possesses a corresponding mirror symmetry as well if the gap function $\Delta(\bk)$ satisfies $\mc{M}_{xy}^{} \Delta(\bk) \mc{M}^{\dagger}_{xy} = \pm \Delta(\bk)$. The matrix Hamiltonian $\mc{H}(\bk)$ of the superconducting state then satisfies
\begin{equation}
\tmc{M}_{xy}^{\pm} \mc{H}(\bk) \left( \tmc{M}^{\pm}_{xy} \right)^{\dagger} = \mc{H}(\bk),
\end{equation}
where
\begin{equation}
\tmc{M}^{\pm}_{xy} = \begin{pmatrix}
\mc{M}^{}_{xy} & 0\\
0 & \pm \mc{M}_{xy}^* \\
\end{pmatrix},
\end{equation}
and $\tmc{M}^{\pm}_{xy}$ is the generalized mirror operator of the superconductor corresponding to the normal state mirror operator $\mc{M}_{xy}$. Since $\mc{H}(\bk)$ commutes with $\tmc{M}^{\pm}_{xy}$, $\mc{H}(\bk)$ is block-diagonal in the basis in which $\tmc{M}^{\pm}_{xy}$ is diagonal, with each block possessing a definite eigenvalue of $\tmc{M}^{\pm}_{xy}$.

We now show that, with an additional constraint on the gap function, individual mirror subsectors of the superconducting state inherit $\mc{C}'$ symmetry when the full Hamiltonian possesses $\mc{C}'$ symmetry. Let the full Hamiltonian $\mc{H}(\bk)$ also possess a mirror symmetry denoted by $\tmc{M}^{\pm}_{xy}$. For now, we will suppress the $\pm$ and write $\tmc{M}_{xy}$. A mirror subsector of $\tmc{M}_{xy}$ possesses $\mc{C}'$ symmetry if and only if a state $|u(\bk) \ra $ and its $\mc{C}'$ conjugate $|u^*(\bk) \ra$ each have $\tmc{M}_{xy}$ eigenvalue $\lambda$.

Therefore, we have
\begin{equation}
\tmc{M}_{xy}|u(\bk) \ra = \lambda | u(\bk) \ra
\end{equation}
and may write
\begin{align}
\begin{split}
\tmc{M}_{xy} |u(\bk) \ra &= \tmc{M}_{xy} \mc{C}' |u^*(\bk)\ra \\
								&= \lambda \mc{C}' |u^*(\bk) \ra \\
								&= \mc{C}' \lambda | u^*(\bk) \ra.
\label{onestarcp}
\end{split}
\end{align}

Enforcing $\mc{C}'$ symmetry in the mirror $\tmc{M}_{xy}$ subsector, we also have
			\begin{equation}
				\tmc{M}_{xy}|u^*(\bk) \ra = \lambda |u^*(\bk) \ra
			\end{equation}
			Then Eq.~\ref{onestarcp} can be written as
			\begin{equation}
				\tmc{M}_{xy}|u(\bk) \ra = \mc{C}' \tmc{M}_{xy}  |u^*(\bk) \ra
				\label{twostarcp}
			\end{equation}
			Using these expressions, we can write
			\begin{align}
			\mc{C}' \tmc{M}_{xy} |u^*(\bk) \ra &= \tmc{M}_{xy} | u(\bk) \ra \\
			\mc{C}' \tmc{M}_{xy}|u^*(\bk) \ra &= \left(\tmc{M}_{xy}^*\tmc{M} \right)\tmc{M}_{xy} | u(\bk) \ra \\
			\mc{C}'\tmc{M}_{xy}|u^*(\bk)\ra &= \lambda^2 \tmc{M}_{xy}^*| u (\bk) \ra \\
			\mc{C}'\tmc{M}_{xy}|u^*(\bk)\ra &= \lambda^2 \tmc{M}_{xy}^* \mc{C}' | u^* (\bk) \ra \label{s1cp}\\
			\mc{C}'\tmc{M}_{xy} &= \lambda^2 \tmc{M}_{xy}^* \mc{C}' \label{s2cp} \\
			\mc{C}' \tmc{M}_{xy} \left(\mc{C}'\right)^{\dagger} &= \lambda^2 \tmc{M}_{xy}^* \label{s3cp}
			\end{align}
			arriving at a constraint for a mirror subsector to inherit $\mc{C}'$ symmetry that is the same as the constraint for a mirror subsector to inherit particle-hole symmetry $\mc{C}$~\cite{ueno2013}.  This constraint is only satisfied for $\tmc{M}_{xy}^-$. Thus, in superconductors with $\mc{C}'$ symmetry and $\tmc{M}_{xy}^-$, the $\mc{C}'$ symmetry of the full Hamiltonian is inherited by each $\tmc{M}_{xy}^-$ subsector.

We can then define the mirror skyrmion number as the skyrmion number computed over the two-dimensional Brillouin zone, but using only occupied states within a single mirror subsector. A topological phase can then be constructed as a system with non-trivial skyrmion number $\mc{Q}$ in one mirror subsector, and non-trivial skyrmion number $-\mc{Q}$ in the other subsector, in analogy to the construction of the quantum spin Hall insulator as a system with a Chern insulator with total Chern number $\mc{C}\neq 0$ in one spin subsector, and a second Chern insulator with total Chern number $-\mc{C}$ in the other spin subsector.

We can generalize these results to three-dimensional systems: the matrix Hamiltonian of the superconducting state then satisfies
\begin{equation}
\tmc{M}^{}_{xy} \mc{H}(k_x, k_y, k_z) \tmc{M}^{\dagger}_{xy} = \mc{H}(k_x, k_y, -k_z),
\end{equation}
and mirror skyrmion numbers can be computed in the $k_x - k_y$ planes where $k_z = 0$ or $\pi$, as is done for mirror Chern numbers~\cite{ueno2013}.

\vspace{2mm}

\textbf{IV. Mirror Chern number and mirror skyrmion number in two dimensions.}

For a Hamiltonian of a two-dimensional system in the $x-y$ plane with mirror symmetry $\mc{M}$ taking $z$ to $-z$, we may block-diagonalize the matrix representation of the Hamiltonian by going to the basis in which the corresponding matrix representation of $\mc{M}$ is diagonal, with each block corresponding to a particular eigenvalue of the matrix representation of $\mc{M}$. The total Chern number $\mc{C}$ for each block---the total mirror Chern number as defined by Ueno \emph{et al.}~\cite{ueno2013}---may then be computed as follows using the method of Fukui \emph{et al.}~\cite{fukui2005, chen2013},
\begin{equation}
\mc{C} = \sum_n c_n,
\end{equation}
where $c_n$ denotes the Chern number of the $n$\textsuperscript{th} band in the block and the sum over $n$ is restricted to bands below the Fermi energy. Here, we evaluate $c_n$ using a discrete lattice corresponding to the discrete values of the crystal momentum in a finite sample with periodic boundary conditions. The Chern number of the $n$\textsuperscript{th} band may then be evaluated as
\begin{equation}
c_n = {1 \over 2 \pi} \sum_{\bk} \mathrm{Im} \ln \left( A^n_{\bk, \hat{x}} A^n_{\bk+\hat{x}, \hat{y}} A^n_{\bk+ \hat{x} + \hat{y}} A^n_{\bk+ \hat{y} , -\hat{y}} \right),
\end{equation}
where $A^n_{\bk, \mu} = \langle n \bk | n \bk + \mu \rangle$ is the Berry connection field and the $\mu$ are the nearest-neighbor vectors of the square momentum-space lattice.

Similarly, we may compute the mirror skyrmion number for each block of the Hamiltonian matrix representation as the skyrmion number $\mc{Q}$ for each block, defined in Eq.~\ref{skyrmnum}.

\textbf{V. Derivation of spin representation in mirror subsectors}

The spin representation is derived from the basis for a given mirror subsector (we focus on mirror subsector $1$ with spin operators and basis given in the main text), as this determines the form of the raising and lowering operators for the spin, which we denote as $\mc{\boldsymbol{S}}_+$ and $\mc{\boldsymbol{S}}_-$, respectively. As particle-hole conjugation and $\mc{C}'$ act within the mirror subsector, the spin representation for mirror subsector $1$ can be decomposed into a normal state component and its particle-hole conjugate. We therefore only discuss here the spin representation for the normal state, which is then used to construct the spin representation for the BdG Hamiltonian of mirror subsector $1$ as given in the main text.

The $\hat{x}$- and $\hat{y}$-components of
 the normal state spin representation are constructed in the standard way from  raising/lowering operators, as $\mc{\boldsymbol{S}}_x = {1 \over 2}\left(\mc{\boldsymbol{S}}_+  + \mc{\boldsymbol{S}}_- \right)$ and $\mc{\boldsymbol{S}}_y = -{i \over 2}\left(\mc{\boldsymbol{S}}_+  - \mc{\boldsymbol{S}}_- \right)$. To construct the raising/lowering operator explicitly and therefore determine $\mc{\boldsymbol{S}}_x$ and $\mc{\boldsymbol{S}}_y$, we first write the representation of the operator which raises a spin $-1/2$ state to spin $+1/2$ irrespective of other degrees of freedom of the normal state in the mirror subsector, which is an orbital degree of freedom. That is, we construct a raising operator that raises spin $-1/2$ in either of the $xz$ or $yz$ orbitals to the spin $+1/2$ state of the $xy$ orbital. The representation for the raising and lowering operators in the basis given in the main text is then
\begin{align}
\mc{\boldsymbol{S}}_+ = \begin{pmatrix}
0 & 1 & 1 \\
0 & 0 & 0 \\
0 & 0 & 0
\end{pmatrix},
\hspace{3mm}
\mc{\boldsymbol{S}}_- = \begin{pmatrix}
0 & 0 & 0 \\
1 & 0 & 0 \\
1 & 0 & 0
\end{pmatrix}
\end{align}

The raising and lowering operators then also include terms for transition from one spin $-1/2$ state in the $yz$ orbital to the other in the $xz$ orbital and vice versa. The raising and lowering operators are then
\begin{align}
\mc{S}_+ = \begin{pmatrix}
0 & 1 & 1 \\
0 & 0 & 1 \\
0 & 0 & 0
\end{pmatrix},
\hspace{3mm}
\mc{S}_- = \begin{pmatrix}
0 & 0 & 0 \\
1 & 0 & 0 \\
1 & 1 & 0
\end{pmatrix}.
\end{align}
This additional transition is counterintuitive, but makes more sense when one considers that it is the $\hat{z}$-component of the spin, which encodes whether a transition corresponds to raising or lowering. To construct the representation of the $\hat{z}$-component of the spin, it is helpful to decompose $\mc{\boldsymbol{S}}_x$ and $\mc{\boldsymbol{S}}_y$ in another way. First, we introduce the following expressions:

 \begin{align}
\sigma_{1,x} &= \begin{pmatrix}
0 & 1 & 0 \\
1 & 0 & 0 \\
0 & 0 & 0 \\
\end{pmatrix} \hspace{1mm}
\sigma_{1,y} = \begin{pmatrix}
0 & -i & 0 \\
i & 0 & 0 \\
0 & 0 & 0 \\
\end{pmatrix} \hspace{1mm}
\sigma_{1,z} = \begin{pmatrix}
1 & 0 & 0 \\
0 & -1 & 0 \\
0 & 0 & 0 \\
\end{pmatrix},
\label{sigma1}
 \end{align}

  \begin{align}
\sigma_{2,x} &= \begin{pmatrix}
0 & 0 & 1 \\
0 & 0 & 0 \\
1 & 0 & 0 \\
\end{pmatrix} \hspace{1mm}
\sigma_{2,y} = \begin{pmatrix}
0 & 0 & -i \\
0 & 0 & 0 \\
i & 0 & 0 \\
\end{pmatrix} \hspace{1mm}
\sigma_{2,z} = \begin{pmatrix}
1 & 0 & 0 \\
0 & 0 & 0 \\
0 & 0 & -1 \\
\end{pmatrix},
\label{sigma2}
 \end{align}

   \begin{align}
\sigma_{3,x} &= \begin{pmatrix}
0 & 0 & 0 \\
0 & 0 & 1 \\
0 & 1 & 0 \\
\end{pmatrix} \hspace{1mm}
\sigma_{3,y} = \begin{pmatrix}
0 & 0 & 0 \\
0 & 0 & -i \\
0 & i & 0 \\
\end{pmatrix} \hspace{1mm}
\sigma_{3,z} = \begin{pmatrix}
0 & 0 & 0 \\
0 & 1 & 0 \\
0 & 0 & -1 \\
\end{pmatrix}.
\label{sigma3}
 \end{align}

We may then rewrite the $\hat{x}$- and $\hat{y}$-components of the spin representation as
 \begin{align}
\mc{S}_{1,x} & \propto \sigma_{1,x} + \sigma_{2,x} + \sigma_{3,x} \\
\mc{S}_{1, y} & \propto \sigma_{1,y} + \sigma_{2,y} + \sigma_{3,y}
 \end{align}
Given that the basis for mirror subsector $1$ is $\Psi_1(\bk) = \{c^{\dagger}_{-\bk, xy, \uparrow}, c^{\dagger}_{-\bk, xz, \downarrow}, c^{\dagger}_{-\bk, yz, \downarrow}, c^{}_{\bk, xy, \uparrow}, c^{}_{\bk, xz, \downarrow}, c^{}_{\bk, yz, \downarrow}\}^{t}$, however, the $\hat{z}$-component of the normal state spin representation for the mirror subsector is instead
 \begin{align}
\mc{S}_{1,z} & \propto \sigma_{1,z} + \sigma_{2,z},
 \end{align}
to account for the redundancy of the spin $-1 / 2$ label. That is, the representation possesses only two complete $\mathrm{SU}(2)$ subalgebras. This furthermore corresponds to generalized commutation relations between the components of the spin representation,
  \begin{align}
 \sum_{j \in \{1,2 \}}\left[ \sigma_{j,\alpha}, \sigma_{j,\beta} \right] =  \sum_{j \in \{1,2 \}}i \varepsilon^{\alpha \beta \kappa} \sigma_{j,\kappa}.
 \end{align}
The commutation relations for spin-$1$ representation satisfy similar generalized commutation relations, which then further simplify to the standard expressions.

\bibliography{p1bib.bib}

\clearpage
\newpage

\end{document}